\documentclass[journal]{IEEEtran}
\usepackage[utf8]{inputenc}
\usepackage[T1]{fontenc}
\usepackage{graphicx}
\usepackage{tabularx,booktabs}
\usepackage{amsmath,amssymb,amsthm,mathtools}
\usepackage{cite}
\usepackage{xcolor}
\usepackage{adjustbox}
\usepackage{gensymb}
\usepackage{bbm}
\usepackage{subcaption} 
\usepackage{listings}
\usepackage{stfloats} 
\usepackage{wasysym} 
\usepackage[margin=0.5in]{geometry}
\usepackage{pgfplots} 
\pgfplotsset{compat=newest}
\usetikzlibrary{patterns}
\usepackage{tikz}

\newtheorem{lemma}{\textbf{Lemma}}

\newtheorem{theorem}{\textbf{Theorem}}

\newtheorem{remark}{\textbf{Remark}}
\DeclareMathOperator\erf{erf}

\title{System-level Analysis of Dual-Mode Networked Sensing: ISAC Integration \& Coordination Gains}

\author{Yasser Nabil, Graduate Student Member, IEEE, Hesham ElSawy, Senior Member, IEEE,\\ and Hossam S. Hassanein, Fellow, IEEE
\thanks{We acknowledge the support of the Natural Sciences and Engineering Research
Council of Canada (NSERC), funding reference numbers RGPIN-2023-03743 and
RGPIN-2025-05001 \\Y.\ Nabil is with the Electrical and Computer Engineering Department, Queen’s University, Kingston, Ontario, Canada. E-mail: \texttt{yasser.nabil@queensu.ca}.\\
H.\ ElSawy and H. S.  Hassanein are with the School of Computing, Queen's University, Kingston, Ontario, Canada. E-mails: \texttt{hesham.elsawy@queensu.ca} and \texttt{ hossam@cs.queensu.ca}.}} 

\begin{document}

\maketitle
\begin{abstract}
This paper characterizes integration and coordination gains in dense millimeter-wave ISAC networks through a dual-mode framework that combines monostatic and multistatic sensing. A comprehensive system-level analysis is conducted, accounting for base station (BS) density, power allocation, antenna misalignment, radar cross-section (RCS) fluctuations, clutter, bistatic geometry, channel fading, and self-interference cancellation (SIC) efficiency.
Using stochastic geometry, coverage probabilities and ergodic rates for sensing and communication are derived, revealing trade-offs among BS density, beamwidth, and power allocation. It is shown that the communication performance sustained reliable operation despite the overlaid sensing functionality. In addition, the results reveal the foundational role of spatial sensing diversity, driven by the dual-mode operation, to compensate for the weak sensing reflections and vulnerability to imperfect SIC along with interference and clutter. To this end, we identify a system transition from monostatic to multistatic-dominant sensing operation as a function of the SIC efficiency. In the latter case, using six multistatic BSs instead of a single bistatic receiver improved sensing coverage probability by over 100\%, highlighting the coordination gain.  Moreover, comparisons with pure communication networks confirm substantial integration gain. Specifically, dual-mode networked sensing with four cooperative BSs can double throughput, while multistatic sensing alone improves throughput by over 50\%.
\end{abstract}


\IEEEpeerreviewmaketitle

\section{Introduction}

The sixth-generation (6G) envisions a unified framework for wireless connectivity and precise sensing, using a single waveform for both sensing and communication (S\&C) \cite{liu2022integrated,zhang2021enabling,cui2024integrated,lu2024integrated,wei2023integrated}. This approach, known as Integrated Sensing and Communications (ISAC) \cite{liu2022integrated,zhang2021enabling}, leverages shared channel characteristics, signal processing, and hardware, enabling base stations (BSs) and mobile users (MUs) to function as sensors. As a result, traditional communication infrastructures are transformed into large-scale radio sensing systems \cite{liu2022integrated,zhang2021enabling,cui2024integrated,lu2024integrated,wei2023integrated}. This supports applications such as vehicular networks, environmental monitoring, indoor services, industrial automation, and drone surveillance, while also facilitating emerging use cases like the metaverse, remote surgery, autonomous driving, and digital twins \cite{liu2022integrated,zhang2021enabling,cui2024integrated,lu2024integrated,wei2023integrated}.
In this context, millimeter-wave (mmWave) frequencies, anticipated to be widely utilized in 6G, are well-suited for ISAC, offering wide bandwidth and narrow beamwidth that enhance communication capacity, radar range, and angular resolutions \cite{liu2022integrated,cui2024integrated}, enabling accurate object localization \cite{liu2022integrated,zhang2021enabling}.

To realize its full potential, ISAC offers two fundamental advantages: integration gain and coordination gain \cite{liu2022integrated,lu2024integrated}. Integration gain results from efficiently utilizing shared wireless resources, which reduces hardware redundancy and enhances spectral and energy efficiency. Coordination gain arises from the mutual reinforcement between S\&C and typically manifests in two forms. The first is sensing-assisted communication, where sensing capabilities improve core communication functions such as handover and beam alignment \cite{10433485}. The second is communication-assisted sensing (also referred to as networked or cooperative sensing), which exploits the connectivity and synchronization inherent in modern communication infrastructure to enable large-scale cooperative sensing across multiple BSs, thereby enhancing coverage, accuracy, and robustness \cite{wang2025cooperative, guo2025integrated,liu2024cooperative,meng2024cooperativenew,armeniakos2025stochastic}.

Achieving these benefits in practical ISAC deployments requires a careful balance between monostatic and multistatic sensing strategies. Monostatic ISAC, with co-located transmitter (Tx) and receiver (Rx), allows compact integration and lower computational overhead but may suffer from strong self-interference (SI) as it relies on full-duplex (FD) operation entailing imperfect self-interference cancellation (SIC) \cite{wei2023integrated,wang2025cooperative,guo2025integrated,liu2024cooperative}. In the absence of FD, multistatic ISAC systems can exploit the cellular infrastructure, with one BS transmitting and other BSs receiving, leveraging spatial diversity and wide angular observations \cite{liu2022integrated,zhang2021enabling,cui2024integrated,wang2025cooperative, guo2025integrated,liu2024cooperative,meng2024cooperativenew,armeniakos2025stochastic}.
However, the reliability of multistatic sensing remains uncertain: longer target-to-BS paths reduce line of sight (LoS) probability, and strong geometry dependence further complicates link quality.

This motivates a dual-mode networked sensing architecture that combines monostatic and multistatic operations to improve sensing accuracy and robustness, even in the presence of environmental blockages.  Such an approach raises critical design questions: should networks invest in advanced FD/SIC capabilities, or instead increase the spatial cooperation level while accepting the associated backhaul overhead for data fusion? Therefore, a rigorous system-level analysis is essential to answer these questions and quantify the gains, trade-offs, and practical implications of networked sensing in dense mmWave ISAC networks.

Beyond single-BS ISAC, a growing body of work studies cooperative or networked sensing. Optimization-centric approaches design coordinated or precoded transmissions to enhance sensing in multistatic ISAC systems while preserving communication \cite{li2023towards,yang2024coordinated,babu2024precoding}. Cell-free architectures utilize distributed apertures for improved sensing via power allocation under communication constraints \cite{behdad2024multi}. Near-field and beamfocusing designs have also been explored for multistatic mmWave ISAC \cite{dehkordi2024multistatic}. In addition, centralized fusion of multiple monostatic returns has been considered \cite{wei2023symbol,li2025distributed}. Despite these advancements, these studies focus on multistatic or monostatic sensing from multiple BSs, without examining their dual operation and interplay. Furthermore, most overlook interference and clutter analysis, except \cite{babu2024precoding} and \cite{behdad2024multi}, which consider interference from a few proximate BSs.

In reality, achieving integration gains in ISAC systems involves overlapping signals in time and frequency, introducing mutual interference and clutter effects \cite{liu2022integrated,zhang2021enabling,wei2023integrated}, which intensify in dense deployments, necessitating large-scale system-level analysis \cite{olson2023coverage}. Moreover, recent ISAC research has shifted from traditional radar metrics to signal-to-interference-plus-noise (SINR)-based evaluations, emphasizing a generalized information-theoretic parameter estimation or inference \cite{olson2023coverage,meng2024network}. Hence, effective ISAC system design requires realistic modeling that incorporates interference, clutter, radar cross-section (RCS) fluctuations, and spatial topology \cite{liu2022integrated,zhang2021enabling,cui2024integrated,lu2024integrated}. 
Stochastic geometry (SG) offers powerful tools for such modeling, enabling large-scale planning to integrate ISAC with cellular networks, particularly in applications demanding wide-area surveillance \cite{meng2024network,olson2023coverage,xu2024performance,sun2024performance,salem2024rethinking,ali2025integrated,10769538,meng2025network}.

Using SG, \cite{olson2023coverage} develops an information-theoretic framework for analyzing coverage and rate in mmWave ISAC networks. Related SG studies balance S\&C via coordinated beamforming with interference nulling \cite{meng2024network}, examine power/spectrum allocation under resolution and rate constraints \cite{xu2024performance}, analyze blockage in urban ISAC \cite{sun2024performance}, and optimize spectral and energy efficiency in dense networks \cite{salem2024rethinking}. While these works provide valuable system-level insight, they focus on monostatic sensing, overlooking the inherently more complex multistatic scenario.
On the other hand, \cite{ali2025integrated} studies a large sub-6\, GHz ISAC network with joint monostatic and bistatic detection, utilizing additional bistatic radars that listen only within the same cell. However, this approach does not fully exploit spatial diversity and also neglects clutter. Moreover, \cite{10769538} provides a system-level analysis of a cooperative ISAC network, deriving the Cramér–Rao Lower Bound (CRLB) for localization accuracy; and \cite{meng2025network} couples coordinated multipoint (CoMP) downlink with multistatic sensing to study antenna-budget distribution across BSs. However, these studies are also sub-6 GHz and overlook sensing interference and clutter effects.

In contrast to prior work, this paper investigates networked sensing in mmWave ISAC networks with a focus on downlink communication and location estimation. We develop a large-scale, topology-aware framework that (i) models both monostatic and multistatic sensing under a unified formulation that captures their interplay, (ii) explicitly accounts for network-wide interference and clutter from all BSs and environmental scatterers, and (iii) jointly evaluates S\&C coverage and rate to reveal system-level trade-offs. To ensure scalability, we adopt a low-overhead selection-combining strategy that harvests spatial diversity without raw-data exchange.
The proposed analysis quantifies the spectral efficiency gains achieved by ISAC relative to conventional communication-only networks operating under comparable resources, thereby demonstrating the integration gain of ISAC. 
Concurrently, we evaluate the enhancement in sensing performance enabled by cooperation among several BSs, inherently connected and synchronized through the underlying cellular network. This improvement demonstrates the coordination gain of ISAC, where the communication infrastructure directly enhances sensing capabilities.
 To this end, the main contributions of this work are summarized as follows:
\begin{itemize}
\item The work identifies ISAC integration and coordination gains in large-scale mmWave ISAC networks through a well-founded mathematical model. 
    \item A novel dual-mode networked sensing approach is explored, integrating monostatic and multistatic sensing using a unified ISAC signal, demonstrating meaningful gains even without FD capabilities.
\item A realistic system-level analysis is conducted, considering interference and clutter from all BSs and environmental scatterers, revealing critical trade-offs between S\&C across key parameters such as power, beamwidth, and BS density.
\item The findings provide informed design guidelines across key system parameters for the effective deployment of large-scale ISAC networks, including conditions under which monostatic or multistatic modes dominate and the performance gains from cooperative sensing versus the associated backhaul overhead required for data fusion.
\end{itemize}
The practical significance of this work lies in demonstrating the resilience of the communication function in the presence of sensing, which can accelerate the adoption of ISAC in mmWave cellular networks. The findings also indicate that FD capability is not a prerequisite for achieving tangible ISAC benefits; instead, standalone multistatic sensing can provide reasonable gains. Additionally, the results quantify the added value of incorporating more cooperative BSs for sensing, enabling service providers to make informed decisions on the optimal number of cooperating BSs needed to balance ISAC performance improvements against backhaul requirements.
Crucially, we identify a SIC-driven mode transition. When residual SIC is strong, the monostatic mode is preferable. Under a limited SIC, the system transitions to multistatic operation, where spatial diversity recovers sensing performance at the expense of increased backhaul overhead.

 The structure of the paper is as follows: Section~\ref{sys_modd} outlines the system model. Section~\ref{ana_synn} details the analysis of the dual-mode networked sensing. Section~\ref{ana_commm} focuses on the communication analysis. Section~\ref{num_ress} discusses the numerical results and design guidelines. Lastly, Section~\ref{con_pp} provides the conclusion.

\section{System Model}\label{sys_modd}

\subsection{ISAC Network and Signal Models}

\begin{figure}[t]
\centering
  \includegraphics[width=0.35\textwidth]{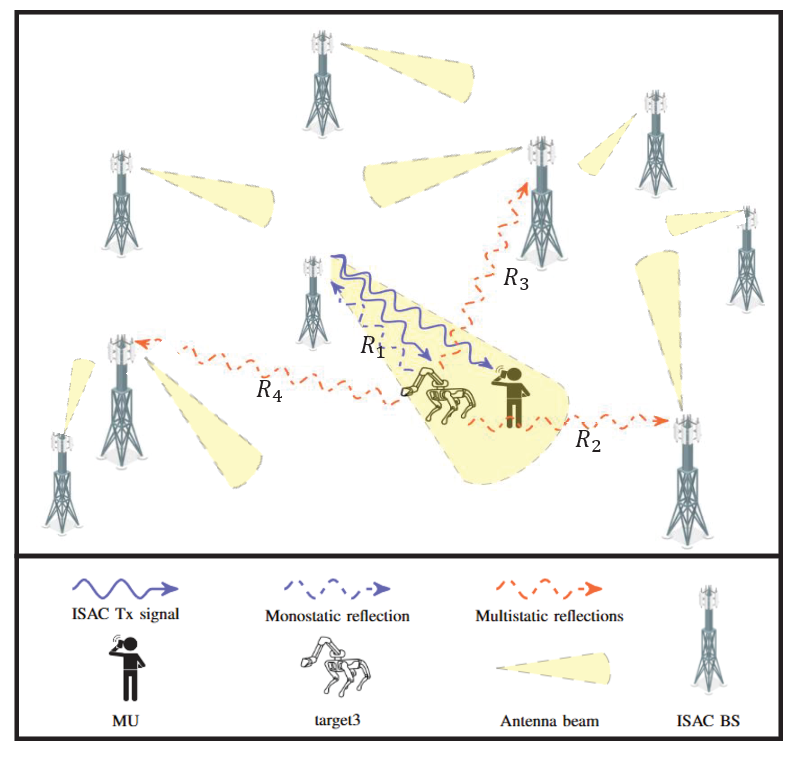}
    \caption{An illustration of the system model with a sensing cluster of four cooperative BSs per target. Each BS forms $M$ beams on $M$ distinct frequency blocks reused across all BSs, with each beam simultaneously sensing a target and serving a MU. 
    For visual clarity, only one beam per BS is shown, and the depicted beams share the same frequency block. The monostatic distance is \( R_1 \), while each \( (R_1, R_n) \) pair forms a bistatic setup, where \( R_n \) is the distance between the target and its \( n^\text{th} \)-nearest BS (\( n \geq 2 \)).
} 
    \label{sys_mod}
\end{figure}

Consider a dense, large-scale mmWave network employing a unified ISAC waveform. The BSs are distributed as a homogeneous Poisson Point Process (PPP) $\boldsymbol{\Phi}_{\mathrm{BS}}$ with intensity $\lambda_{\mathrm{BS}}$, where each BS forms $M$ spatial beams. We model MUs and targets as independent homogeneous PPPs with intensities $\lambda_U$ and $\lambda_T$, respectively. 
 Following \cite{olson2023coverage}, we assume \((\lambda_U,\lambda_T \gg M\lambda_{\mathrm{BS}})\) so each BS can serve a MU and sense a target per beam at any time instant.
Each BS independently transmits its downlink unified ISAC waveforms to serve its MUs and perform monostatic sensing of targets within its cell. To enhance sensing performance, the BSs form cooperative clusters of $N$ nodes.  Within a cluster, a serving BS acts as a monostatic radar for a given target, while the other  $(N-1)$ BSs act as passive receivers, capturing reflections from the same target in a multistatic configuration\footnote{The BSs are interconnected via high-speed 5G fronthaul links (typically optical fiber), enabling reliable downlink multistatic sensing operation and fusion of sensing information \cite{liu2022integrated,zhang2021enabling,cui2024integrated}.}, as illustrated in Fig.~\ref{sys_mod}.

\begin{figure}[t]
\centering
  \includegraphics[width=0.4\textwidth]{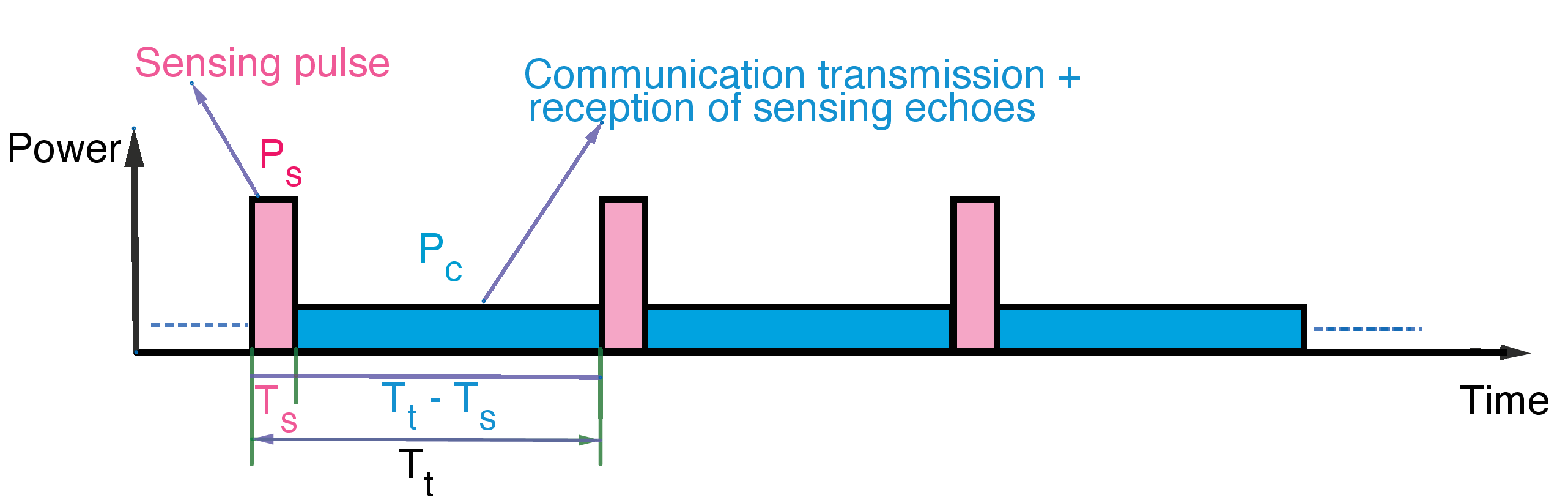}
    \caption{An illustration for the unified ISAC signal.}
    \label{sgn_mod}
\end{figure}

We adopt a low-complexity unified waveform design of a time slot duration \( T_t \) \cite{xiao2022waveform}. The ISAC signal includes a sensing pulse of width \( T_s \), with the remaining time, \( T_t - T_s \), allocated for joint communication and sensing (i.e., reception of sensing echoes) as shown in Fig.~\ref{sgn_mod}. The sensing pulse duration, \( T_s \), is the reciprocal of the unified signal bandwidth \( W_b \), while \( T_t \) is chosen to satisfy the maximum unambiguous range $R_{\text{max}}$, ensuring that the round-trip time from the farthest target does not exceed \( T_t \).
Given a total energy budget \( E_t \) per time slot, a bias factor \( 0 \leq \alpha \leq 1 \) allocates \(\alpha\) of the energy for the sensing pulse and \(( 1-\alpha )\) for communication. Thus, the sensing pulse power is \( P_s = \frac{\alpha E_t}{T_s} \), and the communication power is \( P_c = \frac{(1-\alpha) E_t}{T_t - T_s} \).
This waveform is well-suited for multistatic operations. In monostatic sensing, communication data is transmitted while sensing echoes are received over the same beam, necessitating FD operation with SIC. Given the inherent imperfections of SIC \cite{wei2023integrated, xiao2022waveform}, we define \( \zeta \) as the fraction of power remaining after SIC. Moreover, this work adopts frequency-division duplexing (FDD) with orthogonal uplink/downlink bands and focuses exclusively on downlink ISAC. Consequently, uplink communications do not interfere with sensing reception or downlink transmission. The BS's FD capability is reserved solely for SI suppression on the downlink band.

\begin{remark}
This signal design is a unified signal approach, not a time-sharing scheme, enabling simultaneous S\&C within the same time and frequency. The communication rate is not compromised due to the extremely short sensing pulse relative to the time slot. Nevertheless, it ensures strong sensing performance through a pulse structure with favorable correlation properties and adjustable power to compensate for round-trip path loss.
\end {remark}

\subsection{Beamforming Models}

Consider a multi-beam, multi-frequency antenna system where each beam transmits on a distinct frequency but can receive across multiple frequencies. This setup is feasible with modern beamforming techniques and wideband radio frequency (RF) chains, as demonstrated in \cite{hong2017multibeam,8529214}. To this end, each BS generates \( M \) spatial beams with evenly spaced boresights, providing continuous \( 360^\circ \) coverage, to simultaneously serve and sense \( M \) MUs and \( M \) targets per cell.\footnote{While multiple targets can be sensed per beam if separated by the range resolution, and multiple MUs can be served via user-grouping techniques, this work focuses on a single user and target per beam to reduce analytical complexity. These methods, though enhancing throughput, are left for future work, as the core findings remain valid and extensible to more complex scenarios.}
Each beam operates on a unique frequency block of \( W_b \) Hz to transmit the unified ISAC signal, effectively eliminating intra-cell inter-beam interference. Moreover, universal frequency reuse is applied, ensuring the full bandwidth of \( M W_b \) Hz is utilized across all BSs. The mapping of frequency blocks to beams is random and independent at each BS; consequently, beams using the same frequency block at different BSs point in random directions as illustrated in Fig. \ref{sys_mod}, thereby randomizing interference across the network and preventing correlated interference patterns.
For multistatic operation, signals reflected from targets in neighboring cells randomly arrive at any beam. If a beam receives signals transmitted on any of the $(M-1)$ frequencies other than its own transmission frequency, these signals are useful for multistatic sensing; otherwise, they are treated as interference.

In mmWave networks, the combination of severe path loss, blockage, and narrow beams makes dominant interference events occur primarily when an interferer's transmit mainlobe overlaps with the receiver's mainlobe. Consequently, fine sidelobe ripples have limited impact on spatial coverage statistics, as demonstrated in~\cite{yu2017coverage}. To preserve analytical tractability, we therefore adopt the sidelobe-free cosine model given in (\ref{ant_be_ga}) that explicitly captures these critical TX-RX mainlobe interactions. This approach provides a precise approximation of the mainlobe gain, and closely matches the coverage probability behavior obtained with realistic array patterns, as validated in~\cite{yu2017coverage}.
To this end, each BS beam is modeled using the cosine model given by 
\begin{equation}\label{ant_be_ga}
G(\theta) =
\left\{
	\begin{array}{ll}
		G_m \mathrm{cos}^2 (\frac{d\theta}{2})  &  |\theta| \leq \frac{\pi}{d}\\
		0             &   \text{otherwise}
	\end{array}
\right.,
\end{equation}
where $G_m$ represents the maximum gain, $d$ determines the spread of the beam, and $\theta$ denotes the beam angle relative to the boresight angle. To ensure complete $360^\circ$ coverage, the 3-dB beamwidth of each beam is given by $\theta_B = \frac{\pi}{d} = \frac{2\pi}{M}$, where the number of beams is $M = 2d$. 
Moreover, for simplicity in the analysis, MUs are assumed to use omnidirectional antennas.

\subsection{ Channel and Propagation Models}

Sensing and communication functions occur within a time slot $T_t$, during which channel and sensing parameters are assumed to remain constant but vary across time slots. Considering the dense mmWave setup, we adopt the widely used LoS ball model \cite{yu2017coverage}. That is, we assume that the MU and the target maintain a LoS link with the serving BS responsible for monostatic operation.  However, the multistatic link between the target and other BSs, as well as co-channel interference signals from other BSs, can be either LoS or non-line of sight (NLoS). In mmWave frequencies, according to \cite{akdeniz2014millimeter,rebato2019stochastic}, the probability that a link of distance \( r \) is LoS is given by:
\begin{equation}\label{los_prop}
\boldsymbol{p}_{\boldsymbol{LOS}}(r) = e^{-\gamma r},
\end{equation}
where \(\gamma\) is a parameter determined by the density and geometry of the blockage environment.

To ensure tractability and consistency with widely accepted studies \cite{olson2023coverage,behdad2024multi,yang2024coordinated,salem2024rethinking,dehkordi2024multistatic,10769538}, sensing tasks are considered feasible only when a direct LoS link exists between the target and the BS. This assumption arises because NLoS channels experience excessive delay and require precise knowledge of the reflection geometry. This challenge is intensified in mmWave channels, where significant attenuation causes multipath components to convey much less reliable information.
On the other hand, the communication link and co-channel interference from other BSs are assumed to undergo quasi-static Nakagami-$m$ fading, with fading gains modeled as i.i.d. Gamma random variables (RVs). The shape parameters $m_L$ and $m_N$ correspond to LoS and NLoS conditions, respectively. 
\begin{remark}
The system's cooperative design introduces spatial diversity, ensuring robust sensing. Even if some cooperative BSs have NLoS links to the target, other BSs with LoS links ensure seamless sensing operation.
\end{remark}

The RCS, which quantifies a target's ability to scatter electromagnetic energy back to the radar, significantly influences the strength of the return signal, directly impacting the sensing SINR \cite{barton2004radar,willis2005bistatic}. The monostatic RCS fluctuations of the target, $\sigma_{tm}$, is modeled using the Swerling I model \cite{swerling1960probability}, which follows an exponential probability density function (PDF), given by:
\begin{equation}\label{sw1_rcs}
f(\sigma_{tm})=\frac{1}{\sigma_{\text{av}_t}}\exp{\left(\frac{-\sigma_{tm}}{\sigma_{\text{av}_t}}\right)}, \quad \sigma_{tm} \geq 0,           
\end{equation}
where $\sigma_{\text{av}_t}$ denotes the average RCS of the target. 
By contrast, modeling the bistatic RCS is more challenging, as it depends on the target's capacity to reflect energy from the Tx to the Rx, with the reflection strength heavily influenced by the bistatic angle \( \beta \) \cite{willis2005bistatic}, defined as the interior angle at the target between the TX-target and target-RX lines~\cite{willis2005bistatic}.
To account for the impact of bistatic geometry on the target's RCS, we approximate the bistatic RCS, $\sigma_{tb}$, following the approach in \cite{kell1965derivation} as
\begin{equation}\label{rcs_bi_app}
\sigma_{tb} \approx \sigma_{tm} \cos \left( \frac{\beta}{2} \right).
\end{equation}

This approximation is appropriate for small to moderate bistatic angles when the target surface is smooth and electrically large so that specular reflection dominates~\cite{willis2005bistatic}.
Our cellular ISAC scenario mainly satisfies these conditions: At the mmWave band, typical ground objects such as vehicle body panels or other metallic surfaces are much larger than the wavelength, so electrically large and mostly smooth surfaces. Moreover, in practical cellular deployments, BSs are elevated while targets are near ground level; this geometry keeps the bistatic angle \(\beta\) away from the forward-scatter regime near \(\beta \approx 180^\circ\)~\cite{willis2005bistatic}, where this approximation is inaccurate. Hence, this simple, geometry-aware model captures the essential \(\beta\)-dependence while maintaining the tractability needed for system-level analysis.

From a sensing perspective, clutter emerges as a new form of interference in ISAC networks, caused by reflections from scatterers other than the targets \cite{liu2022integrated,zhang2021enabling,wei2023integrated}. We model clutter scatterers as a homogeneous PPP $\bold\Phi_{cl}$  with intensity $\lambda_{cl}$. The PPP model is adopted for its analytical tractability, serving as a well-established baseline for clutter analysis \cite{chen2012integrated,ram2022estimation}. The PPP captures the variability in both the number and spatial distribution of clutter scatterers across different sensing events, providing a foundation for deriving our spatially averaged performance metrics.\footnote{Clustered point processes can capture environment-specific scatterer aggregation beyond the PPP model.  However, their added analytical complexity, as explained in \cite{armeniakos2025stochastic}, may dilute the paper's main focus. Our PPP-based clutter model provides robust baseline insights and establishes a necessary benchmark for future extensions.}
If the radar resolution cell is defined as the smallest distinguishable unit of space where radar can separate two targets \cite{barton2004radar,willis2005bistatic},  sensing is affected only by clutter from scatterers within the same resolution cell as the target and with comparable RCS. This clutter arrives with nearly the same delay as the target reflection, making it difficult to filter out \cite{shnidman2005radar}.
Moreover, clutter with significantly different RCS is excluded, as it can be mitigated using standard signal processing techniques \cite{rahman2019framework}.

To this end, the monostatic RCS of clutter, \( \sigma_{cm} \), is modeled using the Weibull distribution, which is widely adopted in radar literature  for its mathematical flexibility in representing diverse environmental conditions \cite{sekine1990weibull} with PDF given by:
\begin{equation}\label{clu_wei}
f(\sigma_{cm}) = \frac{k}{\sigma_{\text{av}_{cl}}} \left( \frac{\sigma_{cm}}{\sigma_{\text{av}_{cl}}} \right)^{k-1} \exp \left( - \left( \frac{\sigma_{cm}}{\sigma_{\text{av}_{cl}}} \right)^k \right), \quad \sigma_{cm} \geq 0,     
\end{equation}  
where \( k \) denotes the shape parameter governing the tail behavior of the distribution, and \( \sigma_{\text{av}_{cl}} \) represents the scale parameter. The actual value of \( k \) is typically determined empirically from measured data. For analytical tractability, we assume \( k = 1 \), which reduces the Weibull distribution to the exponential case where \( \sigma_{\text{av}_{cl}} \) precisely equals the average clutter RCS.
Furthermore, since the RCS of relevant clutter is comparable to that of the target, its bistatic counterpart, $\sigma_{cb}$, can be approximated as in (\ref{rcs_bi_app}) such that $\sigma_{cb} \approx \sigma_{cm} \cos \left( \frac{\beta}{2} \right)$.

\section{Dual-mode Networked Sensing Analysis}\label{ana_synn}

In this section, we carry out the dual-mode networked sensing analysis based on a parameter estimation framework grounded in mutual information and SINR statistics. The mutual-information-based approach has gained widespread adoption in ISAC systems, as it extends classical communication metrics such as coverage probability and ergodic rate to sensing applications, thereby enabling a unified performance evaluation framework~\cite{olson2023coverage,meng2024network}. Specifically, mutual information quantifies the amount of information that the received signal conveys about the target parameters of interest. When normalized by the observation time, it yields the sensing information rate, which represents the rate at which target-relevant information is acquired~\cite{olson2023coverage}.

In this context, sensing coverage probability is defined as the probability that the sensing information rate exceeds a predefined threshold, while the sensing ergodic rate denotes the spatial average of this rate over the network ~\cite{olson2023coverage}. These metrics not only offer analytical tractability but also align with fundamental limits on parameter estimation accuracy, such as the CRLB. In particular, the sensing ergodic rate establishes an information-theoretic lower bound on estimation error \cite{liu2023deterministic}. Hence, sensing coverage probability can be interpreted as the fraction of targets whose sensing SINR exceeds a threshold that can be related to a desired estimation-accuracy level.

Mathematically, the sensing coverage probability is formulated as \( \mathbb{P} \left( \text{SINR} > \phi_s \right) \), where $\phi_s$ is a predefined threshold \cite{olson2023coverage}. Here, SINR is the ratio of the received sensing signal power to the combined interference, clutter, and noise, where a higher SINR enhances the accuracy of target parameter estimation \cite{tang2018spectrally}.
Moreover,
as derived in \cite{olson2023coverage,meng2024network}, the sensing information rate is given by
\begin{equation}\label{rat_beg}
\mathcal{R}_s = \mathbb{E}[\log(1 + \text{SINR})].   
\end{equation}

 This network-level perspective fundamentally differs from classical radar metrics: while probability of detection and false-alarm rate address per-link detection performance for specific waveforms and detectors, sensing coverage probability abstracts these details to focus on the fundamental SINR distribution across the entire network.
Crucially, because sensing coverage and rate mirror their communication counterparts, they enable direct comparison and joint design over key system parameters, thereby facilitating the quantification of ISAC gains and identifying design trade-offs in large-scale deployments.

\subsection{Types of Interference and Distance Distributions}
\begin{figure*}[h]
\centering
  \includegraphics[width=0.61\textwidth]{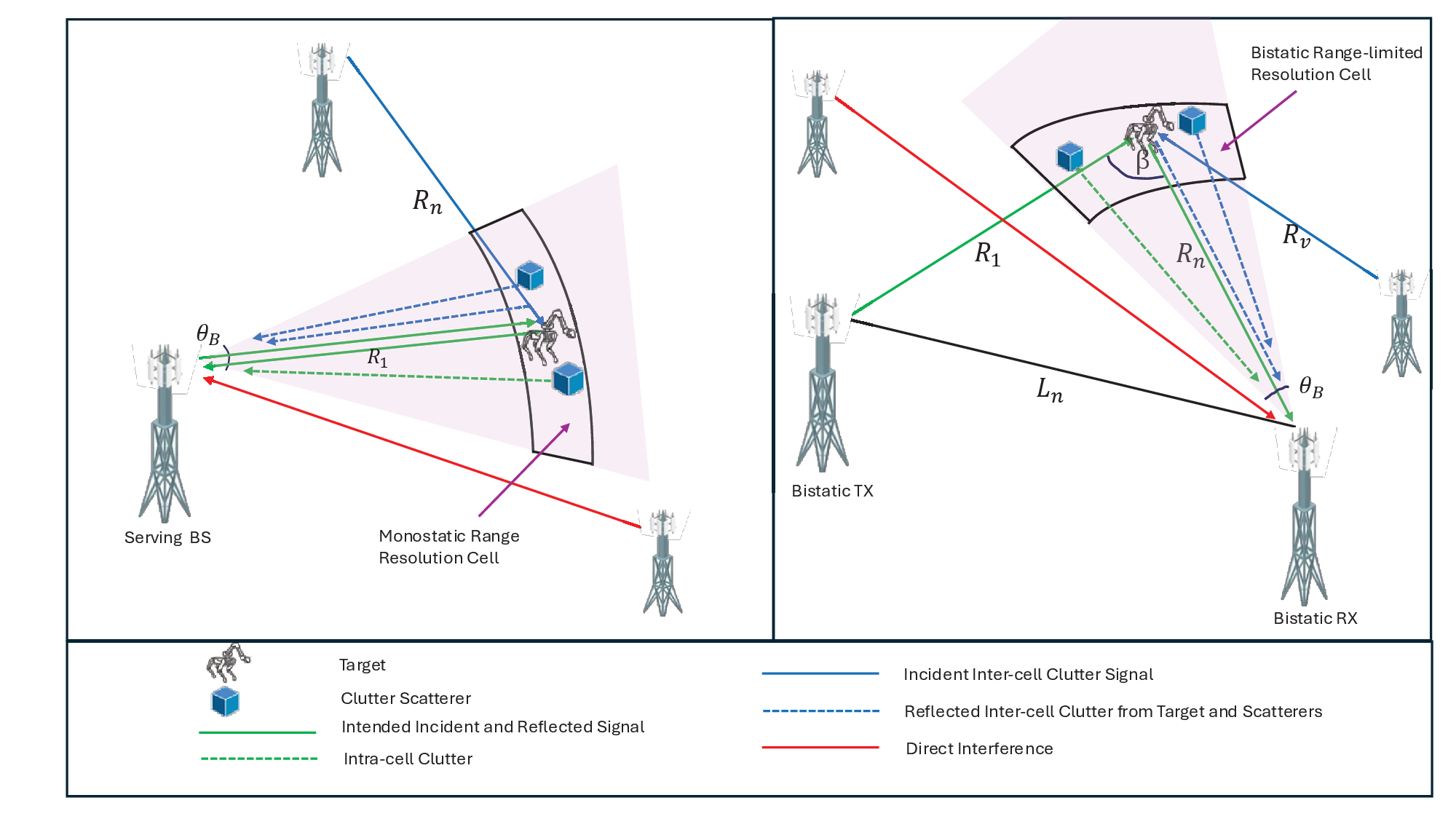}
    \caption{Interference sources in monostatic operation (left) and bistatic operation (right), the bistatic baseline between the TX and RX BSs is indicated by \( L_n \), and $\beta$ is the bistatic angle. Each interference type is shown with a single representative signal for clarity.}
    \label{ill_ntr}
\end{figure*}
 Given the statistical equivalence across all 
$M$ frequencies, we focus on a generic frequency resource used by one beam in every BS. Thus, all BSs in the network are potential interferers.
To this end, the different types of interference for sensing, covering both monostatic and bistatic scenarios, are depicted in Fig. \ref{ill_ntr} and are summarized below:
\begin{enumerate}
    \item \textbf{Direct or co-channel interference from other BSs:} represented by red signals in Fig. \ref{ill_ntr}, these signals do not experience the double path-loss affecting the intended sensing signal. They are analyzed under both LoS and NLoS conditions.
    \item \textbf{Intra-clutter interference:} depicted as dashed green signals in Fig. \ref{ill_ntr}, this arises when the intended signal reflects off scatterers within the target's resolution cell. It is analyzed due to its comparable RCS to the target and a similar delay profile, making it difficult to filter out.
    \item \textbf{Inter-clutter interference:} represented by blue signals in Fig. \ref{ill_ntr}, this interference arises from signals originating from other BSs, reflecting off the target or scatterers within the same resolution cell, and reaching the serving BS. These signals can share similar delay or doppler characteristics with the target's echoes, making them difficult to filter out.
\end{enumerate}

The monostatic sensing of a given target is conducted by its nearest BS.  Based on the properties of PPP, the PDF of \( R_1 \) is:
\begin{equation}\label{ner_dis}
f(R_1) = 2 \pi \lambda_{\mathrm{BS}} R_1 e^{-\pi \lambda_{\mathrm{BS}} R_1^2}, \quad R_1 \geq 0.
\end{equation}
As in \cite{8057736}, the conditional PDF of the distance \( R_n \) to the \( n^{\text{th}} \) nearest BS, given that the distance to the nearest BS is \( R_1 \), is:
\footnotesize
\begin{equation}\label{cond_dstt}
f(R_n \mid R_1) = \frac{2 (\lambda_{\mathrm{BS}} \pi)^{n-1}}{(n-2)!} (R_n^2 - R_1^2)^{n-2} R_n e^{-\lambda_{\mathrm{BS}} \pi (R_n^2 - R_1^2)}, 0 \leq R_1 \leq R_n
\end{equation}
\normalsize
Finally, owing to the structure of the PPP, the angle \( \beta \) between \( R_1 \) and \( R_n \) is uniformly distributed, with the PDF given by:
\begin{equation}\label{bet_distrb}
f(\beta) = \frac{1}{\pi}, \quad 0 \leq \beta \leq \pi.
\end{equation}
\subsection{Monostatic Sensing Analysis}
\begin{figure*}
\hrule
\begin{equation}\label{scnr_mono}
\text{SINR}_M=\frac{\frac{P_s  G_{m}^2 \sigma_{tm}  \lambda^2  }{(4\pi)^3 R_{1}^{2 \eta_L}}}{\sum\limits_{cl\in \bold\Phi_{cl}\cap A_{rm}}\frac{P_s G_{m}^2 \sigma_{cm} \lambda^2 }{(4\pi)^3 R_{1}^{2 \eta_L}}+\sum\limits_{\substack{\text{BS}_{n }\in \bold\Phi_{L_{IC}}\\ n \neq 1}} \frac{P_s  G_{m}^2 \sigma_{tb}  \lambda^2  }{(4\pi)^3 R_1^{\eta_L}  R_n^{\eta_L}}+\sum\limits_{\substack{\text{BS}_{n}\in \bold\Phi_{L_{IC}}\\ n \neq 1}}\sum\limits_{cl\in \bold\Phi_{cl}\cap A_{rm}}\frac{P_s  G_{m}^2 \sigma_{cb}  \lambda^2  }{(4\pi)^3 R_1^{\eta_L}  R_n^{\eta_L}}+K_B T W_b+I_{L_s}+I_{N_s}+ P_c \zeta  }.
\end{equation}
\hrule
\end{figure*}
To evaluate the SINR, we consider a reference BS at the origin without loss of generality. By PPP stationarity \cite{elsawy2016modeling}, the surrounding BS field remains a PPP, which yields the distance distributions discussed previously.
 Following the common practice in SG \cite{elsawy2016modeling}, direct interfering BSs constitute a PPP \(\boldsymbol{\Phi}_{s}\)\footnote{The PPP-based interferer model can approximate other scenarios even if BSs' spatial deployment is not strictly PPP as illustrated in \cite{hmamouche2021new}.} with intensity \(\lambda_{I_s} =   \frac{\lambda_{BS}}{M^2}\), where \(\frac{1}{M^2}\) is the probability of beam alignment between an interfering BS and the reference BS over the same frequency under analysis.
This set \(\boldsymbol{\Phi}_{s}\) is further thinned into two independent PPPs: LoS BSs, \(\boldsymbol{\Phi}_{L_s}\), with intensity \(\boldsymbol{p}_{\boldsymbol{LOS}}(r) \lambda_{I_s}\), and NLoS BSs, \(\boldsymbol{\Phi}_{N_s}\), with intensity \((1 - \boldsymbol{p}_{\boldsymbol{LOS}}(r)) \lambda_{I_s}\). 
Moreover, let \(\boldsymbol{\Phi}_{L_{IC}}\) denote another PPP of BSs generating inter-clutter signals. These BSs contribute to inter-clutter signals if their beams are directed toward the target with LoS paths and use the same frequency block under analysis. The intensity of this PPP is \( \frac{\boldsymbol{p}_{\boldsymbol{LOS}}(r) \lambda_{BS}}{M}\), where \(\frac{1}{M}\) is the probability of a BS beam being directed toward the target.

To estimate location, we consider the monostatic range resolution cell area, defined as \( A_{rm} = \frac{c \, \theta_B \, R_1}{2 \, W_b} \) \cite{barton2004radar}, where \( c \) is the speed of light. 
$ A_{rm}$ is the area of the smallest region where the radar cannot distinguish closely spaced targets, perceiving them as a single target.
Given the typically narrow resolution cell \cite{barton2004radar}, and since only clutter scatterers within the target’s resolution cell are considered, it is reasonable to approximate the range to both the target and clutter as \( R_1 \) for tractability.
Using the radar range equation \cite{barton2004radar}, the monostatic SINR at the typical BS for a target at \( R_1 \) is given in (\ref{scnr_mono}). 

The numerator in (\ref{scnr_mono}) represents the desired echo power, while the denominator represents intra-clutter, inter-clutter reflected from the target, inter-clutter reflected from other scatterers, thermal noise, LoS and NLoS direct interference, and residual SI, respectively.
For simplicity, a constant gain \( G_m \) is assumed within the 3-dB beamwidth, while a gain of zero is assumed outside this range.
In (\ref{scnr_mono}), $cl$ denotes the location of a clutter scatterer, i.e., a point
of the PPP $\boldsymbol{\Phi}_{cl}$.  Hence, the sum $\sum_{cl\in \boldsymbol{\Phi}_{cl}\cap A_{rm}}(\cdot)$ runs over clutter scatterer locations within the resolution cell $A_{rm}$. Moreover, 
\(\eta_L\) and $\eta_N$ are the LoS and NLoS path-loss exponents, and \(\lambda\) is the signal wavelength.  Furthermore, \(K_B\) is the Boltzmann constant, and \(T\) is the system temperature.
\small\(I_{L_s} = \sum\limits_{\substack{\text{BS}_i \in \boldsymbol{\Phi}_{L_s}}} P_c h_{L,i} G_m^2 C_L r_i^{-\eta_L}\)\normalsize and \small\(I_{N_s} = \sum\limits_{\substack{\text{BS}_i \in \boldsymbol{\Phi}_{N_s}}} P_c h_{N,i} G_m^2 C_N r_i^{-\eta_N}\)\normalsize , where \(h_{L,i}\) and \(h_{N,i}\) are the channel gains of the \(i^{\text{th}}\) LoS and NLoS interfering BS, \(C_L\) and \(C_N\) are the LoS and NLoS path-loss intercepts, and  \(r_i\) is the distance between \(\text{BS}_i\) and the reference BS at the origin.
\begin{remark}
Although (\ref{scnr_mono}) represents the monostatic SINR, the bistatic RCS appears in the inter-clutter interference terms. This accounts for a bistatic interference setup, where signals from interfering BSs reflect off the target or scatterers in the same resolution cell and return to the serving BS. This interference strength depends on the geometry, captured by the bistatic RCS.
\end{remark}
Before evaluating the monostatic sensing coverage probability, we first derive the Laplace transform (LT) of the aggregate interference for each interference type. We calculate the direct interference at the reference BS (at the origin), originating from other BSs that are simultaneously transmitting. Accordingly, we define an interference protection region centered at the reference BS. Since no direct interferer can be closer to the serving BS than its nearest neighboring BS, we model the protection region as a disk with radius equal to that nearest-neighbor distance. This distance has the PDF given in (\ref{ner_dis}). The LT of the aggregate direct interference is given in the following lemma.
\begin{lemma} \label{lemma_LT_DMR}
The LT of the aggregate direct interference seen by the typical BS located at the origin is given in (\ref{dir_int}),
\begin{figure*}[h]
\small
\begin{equation}\label{dir_int}
\begin{aligned}
&\mathcal{L}_{\text{direct interference}}(s)= \int_{0}^\infty  \exp \left( - 2\pi \lambda_{BS} \frac{1}{M^2} \int_{R_d}^\infty \bold p_{\bold{LOS}}\left(r\right)  \left( 1 - \left( 1 + \frac{s P_c G_m^2 C_L r^{-\eta_L}}{m_L} \right)^{-m_L} \right) r dr \right)  \\
&\times \exp \left( - 2\pi \lambda_{BS} \frac{1}{M^2} \int_{R_d}^\infty \left( 1 - \bold p_{\bold{LOS}}\left(r\right)  \right) \left( 1 - \left( 1 + \frac{s P_c G_m^2 C_N r^{-\eta_N}}{m_N} \right)^{-m_N} \right) r dr \right) 
\times 2 \pi \lambda_{BS} R_d e^{-\pi \lambda_{BS} R_d^2} dR_d
\end{aligned}
\end{equation}
\normalsize
\hrule
\end{figure*}
where $\bold p_{\bold{LOS}}\left(r\right) $ is given by (\ref{los_prop}).
\begin{IEEEproof}
See Appendix A.
\end{IEEEproof}
\end{lemma}

The LT of intra-clutter interference, arising from scatterers within the target’s resolution cell and with comparable RCS, is presented in the following lemma.
\begin{lemma} \label{intra_clu}
The LT of monostatic intra-clutter interference is
\begin{equation}\label{intra_clu_int}
\begin{aligned}
\mathbb{E}_{cl, \sigma_{cm}} & \left[ \exp\left( - \frac{\phi_s \sum_{cl \in \bold\Phi_{cl} \cap A_{rm}} \sigma_{cm}}{\sigma_{\text{av}_t}} \right) \right] = \\
&\exp \left( - \lambda_{cl} \frac{c \theta_B R_1}{2 W_b} \cdot \frac{\phi_s  \sigma_{\text{av}_{cl}}}{\sigma_{\text{av}_t} + \phi_s  \sigma_{\text{av}_{cl}}} \right)
\end{aligned}
\end{equation}
\begin{IEEEproof}
See Appendix B.
\end{IEEEproof}
\end{lemma}

Inter-clutter interference originates from signals that first illuminate the target or scatterers within same resolution cell and then reflect to the serving BS, forming a two-hop path: \(R_n\) (interfering BS \(\to\) target) and \(R_1\) (target \(\to\) serving BS) as illustrated in Fig.~  \ref{ill_ntr}. Since the target is sensed monostatically by its nearest BS (at distance \(R_1\)), inter-clutter cannot originate from that nearest BS; the closest possible source is the second-nearest BS. Hence, we define an interference protection region as a disk centered at the target with radius equal to the distance from the target to its second-nearest BS, given the nearest BS is at \(R_1\). The PDF of this distance is given in~(\ref{cond_dstt}) with \(n=2\).
  Accordingly, the LT of inter-clutter interference, including reflections from both the target and other scatterers within the same resolution cell, is presented in the following lemma.
\begin{lemma} \label{inter_clu}
The LT of monostatic inter-clutter interference is given in (\ref{inter_clu_int}).
\begin{figure*}[h]
\begin{equation}\label{inter_clu_int}
\small
\begin{aligned}
\mathcal{L}_{\text{inter-clutter}}(s) = & \int_{R_1}^{\infty} \int_{0}^{\pi} 
\exp \left( - 2 \pi \lambda_{BS} \frac{1}{M} \int_{R_{IC}}^{\infty} \bold p_{\bold{LOS}}\left(r\right)  \left( 1 - \frac{1}{1 + s \cos\left( \frac{\beta}{2} \right) r^{-\eta_L} \sigma_{\text{av}_t}} \right) r \, dr \right) \\
& \times \exp \left( - \lambda_{BS} \cdot \frac{1}{M} \int_{R_{IC}}^{\infty}  \bold p_{\bold{LOS}}\left(r\right)  \left( 1 - \exp \left( - \lambda_{cl} \cdot \frac{c \theta_B R_1}{2 W_b} \cdot \frac{s \cos \left( \frac{\beta}{2} \right) r^{-\eta_L} \sigma_{\text{av}_{cl}}}{1 + s \cos \left( \frac{\beta}{2} \right) r^{-\eta_L} \sigma_{\text{av}_{cl}}} \right) \right) 2 \pi r \, dr \right) \\
& \times 2 \lambda_{BS}  R_{IC} e^{-\lambda_{BS} \pi (R_{IC}^2 - R_1^2)}  \,  d\beta  \, dR_{IC}
\end{aligned}
\end{equation}
\normalsize
\hrule
\end{figure*}
\begin{IEEEproof}
See Appendix C.
\end{IEEEproof}
\end{lemma}
It is worth noting that direct interference and inter-clutter interference can be considered uncorrelated because the serving BS and target are typically in different spatial directions from the perspective of interfering BSs. 
Furthermore, intra-clutter interference is independent of direct and inter-clutter interference because the former originates from scatterers illuminated by the serving BS itself, and the latter originates from transmissions of other independent BSs.
Moreover, thermal noise and residual SI are independent of the spatial terms. Therefore, the components of the SINR denominator can be reasonably assumed to be independent. Building on this and the previous lemmas, the monostatic sensing coverage probability is presented in the following theorem.
\begin{theorem} \label{mono_det}
The monostatic sensing coverage probability for a target at a distance \(R_1\) with SINR threshold \(\phi_s\) is provided in (\ref{coV_mon}),
\begin{figure*}[h]
\small
\begin{equation}\label{coV_mon}
\begin{aligned}
 \mathcal{P}_M (R_1) &=  \underbrace{\mathbb{E}_{cl, \sigma_{cm}}\left[ \exp\left( - \frac{\phi_s  \sum_{cl \in \bold\Phi_{cl} \cap A_{rm}} \sigma_{cm}}{\sigma_{\text{av}_t}} \right) \right]}_{\text{Intra-clutter effect}} 
\times \underbrace{\mathcal{L}_{I_{IC1}}\left( \frac{\phi_s  R_1^{\eta_L}}{\sigma_{\text{av}_t}} \right)
\times \mathcal{L}_{I_{IC2}}\left( \frac{\phi_s  R_1^{\eta_L}}{\sigma_{\text{av}_t}} \right)}_{\text{Inter-clutter effect}} 
\times \underbrace{\exp\left( - \frac{\phi_s  (4\pi)^3 R_1^{2 \eta_L} K_B T W_b}{P_s G_m^2 \lambda^2 \sigma_{\text{av}_t}} \right)}_{\text{Noise effect}} \\
&  \times \underbrace{\mathcal{L}_{I_{L_s}}\left( \frac{\phi_s  (4\pi)^3 R_1^{2 \eta_L}}{P_s G_m^2 \lambda^2 \sigma_{\text{av}_t}} \right)
\times \mathcal{L}_{I_{N_s}}\left( \frac{\phi_s  (4\pi)^3 R_1^{2 \eta_L}}{P_s G_m^2 \lambda^2 \sigma_{\text{av}_t}} \right)}_{\text{Direct interference effect}} 
\times \underbrace{\exp\left( - \frac{\phi_s  (4\pi)^3 R_1^{2 \eta_L} P_c \zeta}{P_s G_m^2 \lambda^2 \sigma_{\text{av}_t}} \right)}_{\text{Self-interference effect}}
\end{aligned}
\end{equation}
\normalsize
\hrule
\end{figure*} 
where \(\boldsymbol{p}_{\boldsymbol{LOS}}\left(R_1\right)\) is from (\ref{los_prop}), the direct interference effect from (\ref{dir_int}), the intra-clutter effect from (\ref{intra_clu_int}), and the inter-clutter effect from (\ref{inter_clu_int}).
\begin{IEEEproof}
See Appendix D.
\end{IEEEproof}
\end{theorem}
\subsection{Bistatic Sensing Analysis}
Bistatic sensing involves two BSs: the source BS at a distance \( R_1 \) from the target and the \( n^\text{th} \) cooperative BS at a distance \( R_n \) from the target.
The Rx BS can process the bistatic target return through any of its beams, except for the beam that uses the same frequency resource for its monostatic operation. In this case, the power of the bistatic reflection is too weak to be processed compared to the monostatic echo. Nevertheless, this reflection contributes to inter-clutter interference in the monostatic scenario.
 To this end, the SINR for bistatic sensing at the \( n^\text{th} \)-nearest BS from the target is given in (\ref{scnr_bi}).
\begin{figure*}
\begin{equation}\label{scnr_bi}
\text{SINR}_{B_n}=\frac{\mathbbm{1}_{\{c_1\}}\frac{P_s  G_{m}^2 \sigma_{tb}  \lambda^2  }{(4\pi)^3 R_1^{\eta_L}  R_n^{\eta_L}}}{\sum\limits_{cl\in \bold\Phi_{cl}\cap A_{rb}}\frac{P_s G_{m}^2 \sigma_{cb} \lambda^2 }{(4\pi)^3 R_1^{\eta_L}  R_n^{\eta_L}}+\sum\limits_{\substack{\text{BS}_{v}\in \bold\Phi_{L_{IC}}\\ v \neq 1, v \neq n}}\frac{P_s  G_{m}^2 \sigma_{tb}  \lambda^2  }{(4\pi)^3 R_v^{\eta_L}  R_n^{\eta_L}}+\sum\limits_{\substack{\text{BS}_{v}\in \bold\Phi_{L_{IC}}\\ v \neq 1, v \neq n}}\sum\limits_{cl\in \bold\Phi_{cl}\cap A_{rb}}\frac{P_s  G_{m}^2 \sigma_{cb}  \lambda^2  }{(4\pi)^3 R_v^{\eta_L}  R_n^{\eta_L}}+K_B T W_b+I_{L_s}+I_{N_s} }
\end{equation}
\hrule
\end{figure*}

The numerator represents the desired power received from the target, while the denominator terms correspond to intra-clutter, inter-clutter reflection from the target, inter-clutter reflections from other scatterers, thermal noise, LoS direct interference, and NLoS direct interference. Note that (\ref{scnr_bi}) is free from SI. 
The bistatic range-limited resolution cell area is \( A_{rb} \), and the indicator function \( \mathbbm{1}_{\{\cdot\}} \) equals one if \( \{\cdot\} \) is true and zero otherwise. The condition \( c_1 \) specifies that the target is LoS with the Rx BS, and the bistatic return is received on a different beam than the one using the same frequency, occurring with probability \( \frac{M-1}{M} \). Moreover, \( R_v \) represents the first link distance for inter-clutter interference between an interfering BS and the target (see Fig.~\ref{ill_ntr}).

The bistatic range-limited resolution cell is larger and exhibits a more intricate structure than its monostatic counterpart \cite{willis2005bistatic}. However, for small bistatic baseline \( L_n \) relative to \(( R_1 + R_n )\), typical of dense networks, it is approximated as \cite{willis2005bistatic,ram2022estimation}:
\begin{equation}\label{bis_res_are}
  A_{rb} \approx \frac{c R_n \theta_B}{2 W_b \cos^2\left(\frac{\beta}{2}\right)}.  
\end{equation}
For \(\beta = 0\), this simplifies to the monostatic case, ensuring consistency between both setups.

Following the same approach used in the monostatic analysis, the LT of various interference sources in bistatic sensing is detailed in the following lemmas.
\begin{lemma} \label{intra_clu_bi}
The LT of bistatic intra-clutter interference is:
\begin{equation}\label{intra_clu_int_bi}
\begin{aligned}
\mathbb{E}_{cl, \sigma_{cm}} &\left[ \exp \left( - \frac{\phi_s \sum_{cl \in \bold\Phi_{cl} \cap A_{rb}} \sigma_{cm}}{\sigma_{\text{av}_t}} \right) \right]=\\
& \exp \left( - \lambda_{cl} \frac{c  R_n  \theta_B}{2 W_b\cos^2\left(\frac{\beta}{2}\right)} \cdot \frac{\phi_s \sigma_{\text{av}_{cl}}}{\sigma_{\text{av}_t}+ \phi_s \sigma_{\text{av}_{cl}}} \right) 
\end{aligned}
\end{equation}
\begin{IEEEproof}
Similar to the proof of Lemma \ref{intra_clu}, but using the area of bistatic range-limited resolution cell given by (\ref{bis_res_are}).
\end{IEEEproof}
\end{lemma}

\begin{lemma} \label{inter_clu_b}
The LT of bistatic inter-clutter interference is given in (\ref{inter_clu_int_bi}).
\begin{figure*}[b]
\hrule
\begin{equation}\label{inter_clu_int_bi}
\small
\begin{aligned}
\mathcal{L}_{\text{inter-clutter}}(s) = & \int_{R_1}^{\infty} \int_{0}^{\pi} 
\exp \left( - 2 \pi \lambda_{BS} \frac{1}{M} \int_{R_{IC}}^{\infty} \bold p_{\bold{LOS}}\left(r\right)  \left( 1 - \frac{1}{1 + s \cos\left( \frac{\beta_I}{2} \right) r^{-\eta_L} \sigma_{\text{av}_t}} \right) r \, dr \right) \\
& \times \exp \left( - \lambda_{BS} \cdot \frac{1}{M} \int_{R_{IC}}^{\infty}  \bold p_{\bold{LOS}}\left(r\right)  \left( 1 - \exp \left( - \lambda_{cl} \cdot \frac{c  R_n  \theta_B}{2 W_b\cos^2\left(\frac{\beta}{2}\right)} \cdot \frac{s \cos \left( \frac{\beta_I}{2} \right) r^{-\eta_L} \sigma_{\text{av}_{cl}}}{1 + s \cos \left( \frac{\beta_I}{2} \right) r^{-\eta_L} \sigma_{\text{av}_{cl}}} \right) \right) 2 \pi r \, dr \right) \\
& \times 2 \lambda_{BS}  R_{IC} e^{-\lambda_{BS} \pi (R_{IC}^2 - R_1^2)}  \,  d\beta_I  \, dR_{IC}
\end{aligned}
\end{equation}
\normalsize
\end{figure*}
\begin{IEEEproof}
Similar to the proof of Lemma \ref{inter_clu}, but using the area of bistatic range-limited resolution cell given by (\ref{bis_res_are}).
\end{IEEEproof}
\end{lemma}
Building on the previous lemmas, the bistatic sensing
coverage probability is presented in the following theorem.
\begin{theorem} \label{bi_det}
The bistatic sensing coverage probability at the $n^\text{th}$ nearest BS to the target, where $n \geq 2$, given that the target is at a distance $R_1$ from the source BS and the SINR threshold is $\phi_s$, is expressed in (\ref{bi_cov_pro}),
\begin{figure*}
\hrule
\begin{equation}\label{bi_cov_pro}
\small
\begin{aligned}
\mathcal{P}_{B_n} (R_1) = & \int_{R_1}^{\infty} \int_{0}^{\pi}  \frac{ \bold p_{\bold{LOS}}\left(R_n\right)(M-1)}{M} \times \underbrace{\mathbb{E}_{cl, \sigma_{cm}} \left[ \exp \left( - \frac{\phi_s \sum_{cl \in \bold\Phi_{cl} \cap A_{rb}} \sigma_{cm}}{\sigma_{\text{av}_t}} \right) \right]}_{\text{Intra-clutter effect}} \\
& \times \underbrace{\mathcal{L}_{I_{IC1_b}} \left( \frac{\phi_s R_1^{\eta_L}}{\sigma_{\text{av}_t} \cos \left( \frac{\beta}{2} \right)} \right) \mathcal{L}_{I_{IC2_b}} \left( \frac{\phi_s R_1^{\eta_L}}{\sigma_{\text{av}_t} \cos \left( \frac{\beta}{2} \right)} \right)}_{\text{Inter-clutter effect}} \times \underbrace{\exp \left( - \frac{\phi_s (4\pi)^3 R_1^{\eta_L} R_n^{\eta_L} K_B T W_b}{P_s G_m^2 \lambda^2 \cos \left( \frac{\beta}{2} \right) \sigma_{\text{av}_t}} \right)}_{\text{Noise effect}} \\
& \times \underbrace{\mathcal{L}_{I_{L_s}} \left( \frac{\phi_s (4\pi)^3 R_1^{\eta_L} R_n^{\eta_L}}{P_s G_m^2 \lambda^2 \cos \left( \frac{\beta}{2} \right) \sigma_{\text{av}_t}} \right) \mathcal{L}_{I_{N_s}} \left( \frac{\phi_s (4\pi)^3 R_1^{\eta_L} R_n^{\eta_L}}{P_s G_m^2 \lambda^2 \cos \left( \frac{\beta}{2} \right) \sigma_{\text{av}_t}} \right)}_{\text{Direct interference effect}} \frac{2 (\lambda_{BS} \pi)^{n-1}}{(n-2)!} (R_n^2 - R_1^2)^{n-2} R_n e^{-\lambda_{BS} \pi (R_n^2 - R_1^2)} \cdot \frac{1}{\pi} \, d\beta \, dR_n.
\end{aligned}
\end{equation}
\normalsize
\end{figure*}
where $\boldsymbol p_{\boldsymbol{LOS}}\left(R_n\right)$, direct interference, intra-clutter, and inter-clutter effects are given by (\ref{los_prop}), (\ref{dir_int}), (\ref{intra_clu_int_bi}), and (\ref{inter_clu_int_bi}), respectively.
\begin{IEEEproof}
See Appendix E.
\end{IEEEproof}
\end{theorem}

\subsection{The Cooperative Networked Sensing }
Building on the monostatic and bistatic sensing models, we now analyze dual-mode cooperative sensing, where a target is sensed by its $N$ nearest BSs. In cooperative ISAC, data fusion can occur at two levels \cite{liu2022integrated}: signal-level fusion, where nodes exchange raw echoes for optimal performance at the cost of high backhaul and stringent synchronization \cite{meng2024cooperativenew}, and information-level fusion, where nodes share compact parameter estimates, drastically reducing overhead and synchronization demands \cite{meng2024cooperativenew}.

This work employs a scalable, non-coherent, information-level selection combining strategy. In this approach, no raw data is exchanged. Each cooperating BS computes a single local parameter estimate and an associated SINR-based quality score, sending only a compact metadata report (a few bytes) to a fusion center. The center then selects the estimate from the BS with the highest SINR as the final network output. This captures spatial diversity while minimizing computational complexity, backhaul overhead, and synchronization demands. 

Following this strategy, the networked sensing coverage probability is the probability that the SINR at any of the $N$ cooperative  BSs exceeds a threshold \( \phi_s \), mathematically expressed as:
\begin{equation}\label{joint_fus_form}
  \mathcal{P}_{\text{net}}(R_1) = 1 - \left[(1 - \mathcal{P}_M(R_1)) \prod_{n=2}^{N} (1 - \mathcal{P}_{B_n}(R_1))\right].  
\end{equation}
To evaluate the average cooperative sensing coverage probability, we compute the expectation of (\ref{joint_fus_form}) with respect to \(R_1\), whose PDF is given in (\ref{ner_dis}). 
\begin{remark}\label{ind_asu_net}
(\ref{joint_fus_form}) is derived as the complement of the probability that none of the cooperative BSs' SINRs exceed the threshold \( \phi_s \). This assumes independence among the sensing events at the $N$ BSs, which is well-justified in our system model by several inherent decorrelating factors: (i) the random spatial distribution of BSs according to a PPP creates independent distances and bistatic angles from different BSs to the target; (ii) independent LoS/NLoS blockage states for each link; (iii) uncorrelated interference fields due to random frequency-beam mapping across BSs; and (iv) independent small-scale channel fading gains and RCS fluctuations. While these factors strongly support the independence assumption, we provide comprehensive validation in Section \ref{num_ress} through system-level simulations that capture all spatial correlations.
\end{remark}

To assess the scalability of our approach, we analyze the backhaul overhead of cooperative fusion. 
In our selection combining framework, each cooperative BS transmits a fixed-size payload of $b_{\text{SC}}$ bits per target, for instance, 2 bytes for the estimate and 2 bytes for SINR score, with 16-bit quantization.
With $N$ cooperating BSs, the total payload overhead per target is
$B_{\text{SC}} = N\,b_{\text{SC}}$ bits, which is sufficient to produce a single fused estimate per slot via selection combining.\footnote{The calculations capture the fundamental payload overhead and are independent of transport-layer specifics (e.g., packetization and headers).} 
This modest overhead contrasts sharply with coherent signal-level fusion, which, while representing a theoretical performance upper bound, requires exchanging raw waveform data and incurs orders-of-magnitude higher payload requirements per target \cite{meng2024cooperativenew}.

It is worth highlighting that the main strength of this selection combining approach is its scalability as the network grows.  The payload overhead grows linearly with the number of cooperating BSs ($N$) and with the number of simultaneous sensed targets. Hence, in dense deployments, this approach keeps the backhaul manageable, whereas signal-level fusion can quickly overwhelm the backhaul. This makes it well-suited for scalable, dense mmWave ISAC deployments by minimizing fusion overhead, avoiding stringent phase synchronization, and simplifying processing. Nevertheless, the choice of \(N\) value must balance the performance gains against the linearly increasing overhead, a trade-off that we examine in detail in Section~\ref{num_ress}.

\subsection{Sensing information Rate}

Recalling the mutual information-based sensing rate introduced earlier in this section and defined in (\ref{rat_beg}), we derive the corresponding networked sensing information rate in the following theorem.
\begin{theorem} \label{ses_rate}
The average dual-mode networked sensing information rate obtained through cooperation among \( N \) BSs is:
\begin{equation}
\begin{aligned}
\mathcal{R}_{s_{\text{avg}}}&=M\int_0^\infty\bigg(\int_0^\infty 1 - \bigg[\big(1 - \mathcal{P}_M(R_1,t)\big)\\
& \quad \prod_{n=2}^{N} \big(1 - \mathcal{P}_{B_n}(R_1,t)\big)\bigg] \, dt\bigg)
 2 \pi \lambda_{BS} R_1 e^{-\pi \lambda_{BS} R_1^2} \; dR_1
\end{aligned}
\end{equation}
where $\mathcal{P}_M (R_1,t)$ and $\mathcal{P}_{B_n} (R_1,t)$ can be obtained by replacing $\phi_s$ by $( e^t - 1)$ in Theorem \ref{mono_det} and Theorem \ref{bi_det} respectively. 
\begin{IEEEproof}
Utilizing the integral representation of the expectation of a non-negative RV, the expected value of \(\log(1 + \text{SINR}_{\text{net}}(R_1))\) can be expressed as 
$ \int_0^\infty \mathbb{P}( \text{SINR}_{\text{net}}(R_1) >( e^t - 1)) \, dt$.
Hence, the theorem is proved by utilizing the networked sensing coverage probability expression in (\ref{joint_fus_form}), substituting \(\phi_s\) with \((e^t - 1)\) in Theorems \ref{mono_det} and \ref{bi_det}, and then taking the expectation over \( R_1 \). The scaling factor \( M \) is included because each BS operates \( M \) independent beams, with each beam simultaneously sensing a distinct target.
\end{IEEEproof}
\end{theorem}

\section{Communication Analysis}\label{ana_commm}

This section focuses on analyzing the communication performance of the proposed ISAC system.


\subsection{SINR Formulation}

We start by calculating the SINR at a typical MU, which is assumed to be located at the origin and served by the nearest BS.
For mathematical tractability, interfering BSs are divided into two independent PPPs: 
\(\boldsymbol{\Phi}_{L_c}\) for LOS BSs with intensity \(\boldsymbol{p}_{\boldsymbol{LOS}}(r) \times \lambda_{BS}\), 
and \(\boldsymbol{\Phi}_{N_c}\) for NLOS BSs with intensity \((1 - \boldsymbol{p}_{\boldsymbol{LOS}}(r)) \times \lambda_{BS}\). Moreover,
an interference protection region is defined with a radius equal to the distance between the MU and its serving BS, 
as no other BS can be closer to the MU than the serving BS.
 Hence, the SINR at the typical MU is given as follows:
\begin{equation}\label{sinr_comm}
 \mathrm{SINR_C}=\frac{P_c h_{L,o} G(\theta_m) C_L R_o^{-\eta_L}}{I_{L_c}+I_{N_c}+K_B T W_b}, 
\end{equation}
where \( h_{L,o} \) represents the Gamma-distributed fading gain for the intended channel, and \( R_o \) denotes the distance of the intended link. 
The LOS and NLOS interference are expressed as 
\( I_{L_c} = \sum\limits_{\substack{\text{BS}_i \in \boldsymbol{\Phi}_{L_c}}} P_c h_{L,i} G(\theta_i) C_L r_i^{-\eta_L} \) 
and 
\( I_{N_c} = \sum\limits_{\substack{\text{BS}_i \in \boldsymbol{\Phi}_{N_c}}} P_c h_{N,i} G(\theta_i) C_N r_i^{-\eta_N} \), 
respectively. Here, \( \theta_i \) is the angle between the line connecting the interfering \( \text{BS}_i \) and its intended MU, 
and the line connecting \( \text{BS}_i \) to the reference MU located at the origin.
\begin{remark}
The possible reflections from targets within the ISAC network that reach the MU do exist physically, but they are inherently part of the received multipath and are already captured by the fading channel gains. Hence, the communication SINR in (\ref{sinr_comm}) does not require extra terms for target reflections.
However, the impact of introducing sensing on communication in the unified ISAC signal approach appears in key system parameters such as power, beamwidth, BS density, and shared backhaul capacity, where the requirements for sensing and communication may conflict (i.e., enhancing one can compromise the other). These critical design trade-offs are analyzed in detail in Section \ref{num_ress}.
\end{remark}

\subsection{The Impact of Misalignment Error}

Reducing the beamwidth of BS beams improves sensing and communication. Narrower beams enhance antenna gain, boosting communication coverage and rate. For sensing, they improve angular resolution, enable sensing of more targets, and reduce clutter and interference by shrinking the resolution cell area. However, excessively narrow beams make the system prone to misalignment errors, drastically reducing coverage and rate. Thus, it is crucial to study misalignment effects in ISAC networks and determine an optimal beamwidth that balances improved sensing with reliable communication performance.
In a realistic scenario, a misalignment error angle \(\theta_m\) exists between the BS and the MU, caused by various sources of uncertainty. This angle can be modeled as a zero-mean truncated Gaussian RV \cite{bahadori2019device}, with 
$\theta_m \sim N_t(0, a^2, -\theta_M, \theta_M),$
where \(a^2\) denotes the variance and \(\theta_M\) is the maximum error angle, satisfying \(|\theta_M| < \pi\). To proceed with the analysis, the PDF of the antenna gain \(G(\theta_m)\) must be derived instead of the PDF of \(\theta_m\).

\begin{lemma} \label{lemma_gain_mis}
The PDF of the antenna gain with a truncated Gaussian misalignment error  $\theta_m \sim N_t(0,a^2,- \theta_M ,\theta_M )$  is 
\begin{equation}\label{pdf_gain_ant}
 f\left(G_c\right)= \frac{\sqrt{\frac{2}{d^2 a^2}}\exp{\left(-\frac{2}{d^2 a^2}\;\arccos^2\left(\sqrt{\frac{G_c}{G_m}}\right)\right)}}{\erf{\left(\frac{\theta_M}{\sqrt{2 a^2}}\right)} \sqrt{\pi G_c \left(G_m - G_c\right) }}
 \end{equation}
where 
\small
  \[
  \begin{aligned}[b]
 &G_c \in \left[G_m \; \cos^2 \left(\frac{d\theta_M}{2}\right),G_m\right] \;\;\; \text{for}\;\;\; |\theta_M|< \frac{\pi}{d}\\
&G_c \in \left[0,G_m\right]  \;\;\; \text{for}\;\;\; \frac{\pi}{d} \leq |\theta_M| < \pi 
\end{aligned}
\]
\begin{IEEEproof}
\normalsize
The proof follows starting from the definition of the cumulative distribution function (CDF) of the gain, performing a transformation of a RV based on the relationship between \(\theta_m\) and \(G(\theta_m)\) given in (\ref{ant_be_ga}), substituting the PDF of \(\theta_m\) (a truncated Gaussian), and then taking the derivative.
\end{IEEEproof}
\end{lemma}

\subsection{Communication Coverage Probability}

We begin by calculating the LT of the interference.
\begin{lemma} \label{lemma_LT_comm}
The LTs of the aggregate LOS and NLOS interference seen by the typical MU located at the origin are given by
\begin{equation}\label{LOS_int_com}
\begin{aligned}
\mathcal{L}_{I_{L_c}} (s)&=\exp\left(- \lambda_{BS} \int_{-\frac{\pi}{d}}^{\frac{\pi}{d}}\int_{R_o}^{\infty}\bold p_{\bold{LOS}}\left(r\right) \right.\\
&  \left. \left(1-\left( 1+\frac{s P_c G (\theta_i)C_L r^{-\eta_L}}{m_L}\right)^{-m_L}    \right) r\;dr \;d\theta_i \right)
\end{aligned}
\end{equation}
\begin{equation}\label{NLOS_int_com}
\begin{aligned}
\mathcal{L}_{I_{N_c}} (s)&=\exp\left(- \lambda_{BS} \int_{-\frac{\pi}{d}}^{\frac{\pi}{d}}\int_{R_o}^{\infty}\left(1-\bold p_{\bold{LOS}}\left(r\right)\right) \right.\\
& \left. \left(1-\left( 1+\frac{s P_c G (\theta_i)C_N r^{-\eta_N}}{m_N}\right)^{-m_N}    \right) r\;dr \;d\theta_i \right)
\end{aligned}
\end{equation}
where $\bold p_{\bold{LOS}}\left(r\right)$  is given in (\ref{los_prop}).
\begin{IEEEproof}
See Appendix F.
\end{IEEEproof}
\end{lemma}
Utilizing the results of Lemma \ref{lemma_gain_mis} and Lemma \ref{lemma_LT_comm}, the coverage probability is derived in the following theorem.

\begin{theorem} \label{com_cov_prob}
The average communication coverage probability with a predefined SINR threshold $\phi_c$ is given as follows:
\begin{equation}
\begin{aligned}[b]
\mathcal{P}_{\text{com}}&= \int_{0}^{\infty} 2 \pi \lambda_{BS} R_o e^{-\pi \lambda_{BS} R_o^2} \sum_{n=1}^{m_L} \left(-1\right)^{n+1} {m_L \choose n}\\
&\int \exp\left(-\;\frac{k_L\;n \;\phi_c \; R_o^{\eta_L}\; K_B T W_b}{P_c G_c C_L}\right) \mathcal{L}_{I_{L_c}} \left(\frac{k_L\;n \;\phi_c \; R_o^{\eta_L}}{P_c G_c C_L}\right)\\
&\mathcal{L}_{I_{N_c}} \left(\frac{k_L\;n \;\phi_c \; R_o^{\eta_L}}{P_c G_c C_L}\right)f\left(G_c\right) dG_c\; dR_o.
\end{aligned}
\end{equation}
where \(k_L = m_L(m_L!)^{-\frac{1}{m_L}}\), while \(\mathcal{L}_{I_{L_c}}\) and \(\mathcal{L}_{I_{N_c}}\) are defined in (\ref{LOS_int_com}) and (\ref{NLOS_int_com}), respectively. Additionally, \(f(G_c)\) and its integration limits are provided in Lemma \ref{lemma_gain_mis}.
\begin{IEEEproof}
See Appendix G.
\end{IEEEproof}
\end{theorem}

\subsection{Communication Rate}

Similar to calculating the sensing rate, the average communication rate per cell is calculated in the following theorem.
\begin{theorem} \label{comm_rate}
The average communication  rate is given as follows:
\small
\begin{equation}
\mathcal{R}_{c_{\text{avg}}}=\frac{M(T_t - T_s)}{T_t} \int_0^\infty \mathcal{P}_{\text{com}}(t)\; dt 
\end{equation}
\normalsize
where $\mathcal{P}_{\text{com}}(t)$ is given by replacing $\phi_c$ by $( e^t - 1)$ in Theorem \ref{com_cov_prob}, where the term \(\frac{T_t - T_s}{T_t}\) is to account for the actual communication duration, and the multiplication by $M$ since a BS is serving $M$ users over $M$ beams simultaneously.
\begin{IEEEproof}
Similar to the proof of Theorem \ref{ses_rate} 
\end{IEEEproof}
\end{theorem}

\section{Numerical Results and Design Guidelines}\label{num_ress} 

We begin by validating our analytical framework through Monte Carlo simulations and investigating the impact of critical system parameters. Subsequently, we analyze the fundamental trade-offs between S\&C functions to establish design guidelines for practical ISAC network deployment.

\subsection{Numerical and Simulation Results}

The Monte Carlo simulation preserves exact spatial correlations by jointly evaluating sensing SINRs under common network realization, without invoking any independence assumptions. Each round proceeds as follows.
Two independent PPPs are generated for BSs and clutter scatterers. A typical MU is randomly placed and associated with its nearest BS, while a typical target is randomly placed and sensed by its $N$ nearest BSs. The nearest BS performs monostatic sensing, with the remaining $N-1$ providing multistatic sensing.
We compute actual distances between the target and cooperative BSs, along with corresponding bistatic angles. RCSs and fading gains are sampled from their respective distributions. Interference is calculated based on actual geometry: direct interference arises from BSs aligned toward the serving BS, and inter-clutter interference from BSs oriented toward the target with a LoS path. Each clutter scatterer uses its actual distance to the serving BS, not the \(R_1\) approximation.
The communication SINR is evaluated at the MU, while the sensing SINRs are computed at the $N$ cooperative BSs, with the highest among them selected as the sensing SINR for this realization. Coverage probabilities and rates are averaged over $2 \times 10^5$ runs. The simulation area is 25 km$^2$, and the operating frequency lies within the 28 GHz band.

Moreover, we set \(T_t=\tfrac{2R_{\max}}{c}\) with \(R_{\max}=3R_{\text{eff}}\)~\cite{olson2023coverage}, where \(R_{\text{eff}}=\sqrt{1/(\pi\lambda_{\text{BS}})}\) is the effective Voronoi-cell radius, and the sensing pulse duration is \(T_s=\tfrac{1}{W_b}\). 
Unless stated otherwise, all numerical values are taken from Table~\ref{tab:system_parameters}.
It is worth highlighting that the parameter values in Table 1 are selected based on established literature to represent practical mmWave ISAC scenarios. Nevertheless, our analytical framework provides general insights that are not tied to specific parameter values. The fundamental trade-offs and design guidelines remain applicable across a wide range of realistic system configurations.

\begin{table} 
  \begin{center}
     \caption{ Numerical Parameters}
    \begin{tabular}{l@{}cl}
      \hline
      \textbf{Parameter Description} & \textbf{Symbol} & \textbf{Value} \\
      \hline
Number of antenna beams         & $M$                     & $12$                            \\ 
3-dB beamwidth and   spread parameter            & $\theta_B$,  $d$           & $30^\circ $, 6 \\ 
 Antenna beam maximum gain               & $G_m$                   & $10 \, \text{dBi}$              \\ 
Path loss intercepts                  & $C_L$, $C_N$                      & $-61.4 \, \text{dB}$, $-72 \, \text{dB}$     \cite{yu2017coverage}        \\ 
Path loss exponents       & $\eta_L$   , $\eta_N$                & $2$ , $4$           \cite{yu2017coverage}                     \\ 
Nakagami fading parameters   & $m_L$, $m_N$                   & $3$, $2$     \cite{yu2017coverage}                          \\ 
LoS blockage parameter        & $\gamma$                & $0.0149$       \cite{rebato2019stochastic}                  \\ 
Bandwidth           & $W_b$                      & $208$ MHz    \cite{ram2022estimation}                   \\ 
Clutter density                 & $\lambda_{\text{cl}}$   & $0.01\; \text{m}^{-2}$          \cite{ram2022estimation}              \\ 
Number of collaborative BSs       & $N$    &  4\\
Average RCS of target and clutter           & $\sigma_{\text{av}_t}$, $\sigma_{\text{av}_{cl}}$   & $1 \; \text{m}^2$      \cite{xiao2022waveform,ram2022estimation}                       \\ 
Transmission powers & $P_s$, $P_c$                   & $0.9 \, \text{W}$, $0.1 \, \text{W}$     \cite{xiao2022waveform}           \\ 
Temperature                     & $T$                     & $300 \, \text{K}$               \\ 
Fraction of power remaining after SIC  & $\zeta$                 & $10^{-12}$    \cite{ali2024successive}                \\ 
 Base station density            & $\lambda_{\text{BS}}$   & $250 \, \text{BS/km}^2$       \\ 
SINR thresholds            &   $\phi_s$,  $\phi_c$                       &     0 dB                           \\
Variance and maximum  error angle &  $a^2$, $\theta_M$  & 1, 0.2 $\pi$    \cite{bahadori2019device} \\ 
 \end{tabular}
\label{tab:system_parameters}
  \end{center}
\end{table}

\begin{figure*}
\centering
\subfloat[\label{eff_sources_mono}]{%
%
%
\definecolor{mycolor1}{rgb}{0.00000,0.44700,0.74100}%
\definecolor{mycolor2}{rgb}{0.85000,0.32500,0.09800}%
\definecolor{mycolor3}{rgb}{0.92900,0.69400,0.12500}%
\definecolor{mycolor4}{rgb}{0.49400,0.18400,0.55600}%
\definecolor{mycolor5}{rgb}{0.46600,0.67400,0.18800}%
\definecolor{mycolor6}{rgb}{0.30100,0.74500,0.93300}%
\definecolor{mycolor7}{rgb}{0.63500,0.07800,0.18400}%
\begin{tikzpicture}[scale=0.42, transform shape,font=\Large]

\begin{axis}[%
width=4.521in,
height=3.566in,
at={(0.758in,0.481in)},
scale only axis,
xmin=-10,
xmax=10,
xlabel style={font=\Large\color{white!15!black}},
xlabel={SINR Threshold (dB)},
ymin=0.1,
ymax=1,
ylabel style={font=\Large\color{white!15!black}},
ylabel={Sensing Coverage Probability},
axis background/.style={fill=white},
xmajorgrids,
ymajorgrids,
legend style={at={(0.003,0.451)}, anchor=south west, legend cell align=left, align=left, draw=white!15!black,font=\Large}
]

\addplot [line width=0.6mm, color=mycolor1]
  table[row sep=crcr]{%
-10	0.9685\\
-9	0.9659\\
-8	0.9631\\
-7	0.9604\\
-6	0.9575\\
-5	0.9544\\
-4	0.9511\\
-3	0.9474\\
-2	0.9433\\
-1	0.9385\\
0	0.9328\\
1	0.926\\
2	0.9176\\
3	0.9074\\
4	0.8949\\
5	0.8794\\
6	0.8604\\
7	0.8371\\
8	0.8087\\
9	0.7743\\
10	0.7331\\
};
\addlegendentry{Direct interference + noise}

\addplot [line width=0.6mm, color=mycolor2]
  table[row sep=crcr]{%
-10	0.9619\\
-9	0.9577\\
-8	0.9533\\
-7	0.9484\\
-6	0.9431\\
-5	0.9373\\
-4	0.9309\\
-3	0.9239\\
-2	0.9162\\
-1	0.9077\\
0	0.8983\\
1	0.8879\\
2	0.8762\\
3	0.863\\
4	0.8479\\
5	0.8305\\
6	0.8101\\
7	0.7861\\
8	0.7577\\
9	0.7241\\
10	0.6845\\
};
\addlegendentry{Direct + intra-clutter + noise}

\addplot [line width=0.6mm, color=mycolor3]
  table[row sep=crcr]{%
-10	0.959791054674701\\
-9	0.955125538594571\\
-8	0.949991909358305\\
-7	0.944315221223317\\
-6	0.938007535562311\\
-5	0.930968909687832\\
-4	0.923088458254987\\
-3	0.914244490588864\\
-2	0.90430246948666\\
-1	0.893109673861297\\
0	0.880486134489199\\
1	0.866212568993181\\
2	0.850017310235276\\
3	0.831565112369267\\
4	0.810450928554432\\
5	0.786201382795809\\
6	0.758286108861261\\
7	0.726141121416286\\
8	0.689206583825634\\
9	0.646982585760481\\
10	0.599106987979931\\
};
\addlegendentry{Direct + intra-clutter + inter-clutter + noise}

\addplot [color=white, draw=none, mark=x, thick, mark size=4pt, mark options={ black}]
  table[row sep=crcr]{%
-10	0.958228\\
-9	0.953504\\
-8	0.948128\\
-7	0.942136\\
-6	0.935652\\
-5	0.928056\\
-4	0.9197\\
-3	0.910496\\
-2	0.900072\\
-1	0.888072\\
0	0.87474\\
1	0.859728\\
2	0.842596\\
3	0.823808\\
4	0.801844\\
5	0.776804\\
6	0.748688\\
7	0.716916\\
8	0.67974\\
9	0.638032\\
10	0.590092\\
};
\addlegendentry{Simulations}

\addplot [line width=0.6mm, color=mycolor1]
  table[row sep=crcr]{%
-10	0.4889\\
-9	0.4849\\
-8	0.4803\\
-7	0.4749\\
-6	0.4687\\
-5	0.4614\\
-4	0.453\\
-3	0.4432\\
-2	0.4318\\
-1	0.4187\\
0	0.4037\\
1	0.3866\\
2	0.3673\\
3	0.3457\\
4	0.322\\
5	0.2961\\
6	0.2683\\
7	0.2391\\
8	0.2091\\
9	0.1788\\
10	0.1492\\
};

\addplot [line width=0.6mm, color=mycolor2]
  table[row sep=crcr]{%
-10	0.4364\\
-9	0.4277\\
-8	0.4183\\
-7	0.4081\\
-6	0.3972\\
-5	0.3856\\
-4	0.3732\\
-3	0.3601\\
-2	0.3464\\
-1	0.332\\
0	0.317\\
1	0.3013\\
2	0.2848\\
3	0.2675\\
4	0.2492\\
5	0.23\\
6	0.2096\\
7	0.1883\\
8	0.1663\\
9	0.1438\\
10	0.1214\\
};

\addplot [line width=0.6mm, color=mycolor3]
  table[row sep=crcr]{%
-10	0.433\\
-9	0.424\\
-8	0.4142\\
-7	0.4036\\
-6	0.3922\\
-5	0.38\\
-4	0.3669\\
-3	0.3531\\
-2	0.3386\\
-1	0.3233\\
0	0.3072\\
1	0.2905\\
2	0.2729\\
3	0.2545\\
4	0.2351\\
5	0.2149\\
6	0.1939\\
7	0.1721\\
8	0.15\\
9	0.1279\\
10	0.1064\\
};

\addplot [color=white, draw=none, mark=x, thick, mark size=4pt, mark options={ black}]
  table[row sep=crcr]{%
-10	0.423874\\
-9	0.41536\\
-8	0.406236\\
-7	0.395968\\
-6	0.385042\\
-5	0.373434\\
-4	0.361072\\
-3	0.347504\\
-2	0.333538\\
-1	0.318704\\
0	0.30309\\
1	0.28671\\
2	0.269696\\
3	0.251656\\
4	0.23249\\
5	0.212282\\
6	0.191778\\
7	0.170242\\
8	0.148742\\
9	0.126664\\
10	0.105394\\
};

\definecolor{darkgreen}{rgb}{0.00000,0.39200,0.00000} 

\draw [stealth-,thick, darkgreen] (axis cs:-1.8,0.92) -- (axis cs:-1.5,0.83) node[below]{Monostatic};
\draw [thick, darkgreen] (-2,0.92) ellipse (0.5cm and 0.5cm); 

\draw [stealth-,thick, darkgreen] (axis cs:-4.5,0.42) -- (axis cs:-3.8,0.22) node[below]{Bistatic}; 
\draw [thick, darkgreen] (axis cs:-4.5,0.42) ellipse (0.7cm and 0.7cm); 

\end{axis}

\begin{axis}[%
width=5.833in,
height=4.375in,
at={(0in,0in)},
scale only axis,
xmin=0,
xmax=1,
ymin=0,
ymax=1,
axis line style={draw=none},
ticks=none,
axis x line*=bottom,
axis y line*=left
]
\end{axis}
\end{tikzpicture}%
}%
\subfloat[ \label{dist_effect}]{%
%
%

\definecolor{mycolor1}{rgb}{0.00000,0.44700,0.74100}%
\definecolor{mycolor2}{rgb}{0.85000,0.32500,0.09800}%
\definecolor{mycolor3}{rgb}{0.92900,0.69400,0.12500}%
\definecolor{mycolor4}{rgb}{0.49400,0.18400,0.55600}%

\begin{tikzpicture}[scale=0.42, transform shape,font=\Large]

\begin{axis}[%
width=4.521in,
height=3.566in,
at={(0.758in,0.481in)},
scale only axis,
xmin=-10,
xmax=5,
xlabel style={font=\Large\color{white!15!black}},
xlabel={SINR Threshold (dB)},
ylabel style={font=\Large\color{white!15!black}},
ylabel={Sensing Coverage Probability},
ymin=0.3,
ymax=1,
axis background/.style={fill=white},
xmajorgrids,
ymajorgrids,
legend style={at={(0.05,0.1)}, anchor=south west, legend cell align=left, align=left, draw=white!15!black}
]

\addplot [line width=0.6mm, color=mycolor1] table[row sep=crcr]{%
-10 0.99706\\ -9 0.99638\\ -8 0.99564\\ -7 0.99448\\ -6 0.99308\\ -5 0.9915\\ -4 0.98988\\ -3 0.98778\\ -2 0.98594\\ -1 0.9834\\ 0 0.98098\\ 1 0.97814\\ 2 0.97518\\ 3 0.97222\\ 4 0.96892\\ 5 0.965\\ 6 0.96098\\ 7 0.95676\\ 8 0.9523\\ 9 0.94752\\ 10 0.94226\\};
\addlegendentry{Monostatic}

\addplot [line width=0.6mm, color=mycolor2] table[row sep=crcr]{%
-10 0.9986\\ -9 0.99838\\ -8 0.99798\\ -7 0.99732\\ -6 0.99654\\ -5 0.99574\\ -4 0.99472\\ -3 0.99336\\ -2 0.99224\\ -1 0.99082\\ 0 0.98936\\ 1 0.98762\\ 2 0.9857\\ 3 0.98372\\ 4 0.98134\\ 5 0.97872\\ 6 0.97598\\ 7 0.97296\\ 8 0.97008\\ 9 0.96656\\ 10 0.96242\\};
\addlegendentry{Dual-mode Networked Sensing 2}

\addplot [line width=0.6mm, color=mycolor3] table[row sep=crcr]{%
-10 0.99912\\ -9 0.99896\\ -8 0.99876\\ -7 0.99834\\ -6 0.99774\\ -5 0.99714\\ -4 0.9963\\ -3 0.99538\\ -2 0.99456\\ -1 0.99346\\ 0 0.99228\\ 1 0.99088\\ 2 0.9895\\ 3 0.98782\\ 4 0.9857\\ 5 0.98362\\ 6 0.98144\\ 7 0.97926\\ 8 0.97694\\ 9 0.97408\\ 10 0.97042\\};
\addlegendentry{Dual-mode Networked Sensing 3}

\addplot [line width=0.6mm, color=mycolor4] table[row sep=crcr]{%
-10 0.99932\\ -9 0.99918\\ -8 0.99904\\ -7 0.9987\\ -6 0.99826\\ -5 0.99784\\ -4 0.99726\\ -3 0.99652\\ -2 0.99578\\ -1 0.99488\\ 0 0.99386\\ 1 0.99264\\ 2 0.99134\\ 3 0.98992\\ 4 0.98818\\ 5 0.98636\\ 6 0.98456\\ 7 0.98276\\ 8 0.98068\\ 9 0.97796\\ 10 0.97456\\};
\addlegendentry{Dual-mode Networked Sensing 4}

\addplot [color=white, draw=none, mark=x, thick, mark size=4pt, mark options={ black}]
  table[row sep=crcr]{%
-10	0.91378\\
-9	0.90314\\
-8	0.890124\\
-7	0.874372\\
-6	0.85612\\
-5	0.833564\\
-4	0.8063\\
-3	0.774604\\
-2	0.737644\\
-1	0.695396\\
0	0.64672\\
1	0.592036\\
2	0.531628\\
3	0.465664\\
4	0.395408\\
5	0.322844\\
6	0.251912\\
7	0.184876\\
8	0.126332\\
9	0.078984\\
10	0.043572\\
};
\addlegendentry{Simulations}

\addplot [color=white, draw=none, mark=x, thick, mark size=4pt, mark options={ black}]
  table[row sep=crcr]{%
-10	0.94398\\
-9	0.935876\\
-8	0.925908\\
-7	0.913264\\
-6	0.89834\\
-5	0.879192\\
-4	0.855596\\
-3	0.827456\\
-2	0.793344\\
-1	0.75278\\
0	0.704672\\
1	0.649412\\
2	0.5864\\
3	0.5156\\
4	0.43884\\
5	0.358128\\
6	0.278616\\
7	0.203292\\
8	0.137924\\
9	0.085496\\
10	0.0468\\
};

\addplot [color=white, draw=none, mark=x, thick, mark size=4pt, mark options={ black}]
  table[row sep=crcr]{%
-10	0.959216\\
-9	0.952592\\
-8	0.944324\\
-7	0.933648\\
-6	0.920496\\
-5	0.9033\\
-4	0.881868\\
-3	0.85556\\
-2	0.822792\\
-1	0.782788\\
0	0.734236\\
1	0.677832\\
2	0.612684\\
3	0.538844\\
4	0.45792\\
5	0.373\\
6	0.289152\\
7	0.21\\
8	0.141824\\
9	0.087392\\
10	0.047596\\
};

\addplot [color=white, draw=none, mark=x, thick, mark size=4pt, mark options={ black}]
  table[row sep=crcr]{%
-10	0.968044\\
-9	0.962368\\
-8	0.955056\\
-7	0.945556\\
-6	0.9336\\
-5	0.917548\\
-4	0.897504\\
-3	0.87214\\
-2	0.839824\\
-1	0.800312\\
0	0.751252\\
1	0.693408\\
2	0.626424\\
3	0.550408\\
4	0.466792\\
5	0.379376\\
6	0.293248\\
7	0.212388\\
8	0.143008\\
9	0.087956\\
10	0.047812\\
};

\addplot [line width=0.6mm, color=mycolor1] table[row sep=crcr]{%
-10	0.916789928976338\\
-9	0.906264795607331\\
-8	0.893583147325326\\
-7	0.878293022385579\\
-6	0.859895663714237\\
-5	0.837854307123956\\
-4	0.811607831996506\\
-3	0.780590333998053\\
-2	0.744258560275827\\
-1	0.702130948640062\\
0	0.653844607773\\
1	0.599238525700077\\
2	0.538470581655371\\
3	0.472168685429716\\
4	0.401599874191573\\
5	0.328814193316785\\
6	0.256687083742824\\
7	0.188759082165411\\
8	0.128783095869016\\
9	0.079973983139387\\
10	0.044128961777972\\
};

\addplot [line width=0.6mm, color=mycolor2] table[row sep=crcr]{%
-10	0.945673808830072\\
-9	0.937708832332078\\
-8	0.927881724609784\\
-7	0.915706234132162\\
-6	0.900659300565252\\
-5	0.882102569544285\\
-4	0.859274064776078\\
-3	0.831477579014553\\
-2	0.797802634027998\\
-1	0.757631101765651\\
0	0.709935627529463\\
1	0.654468265996298\\
2	0.591142164994533\\
3	0.520148551924155\\
4	0.443372990174008\\
5	0.36306883214667\\
6	0.283018200488318\\
7	0.207093126908473\\
8	0.140436493178672\\
9	0.086620251085188\\
10	0.047405687696997\\
};

\addplot [line width=0.6mm, color=mycolor3] table[row sep=crcr]{%
-10	0.960793222442192\\
-9	0.954272302979649\\
-8	0.946097358607845\\
-7	0.935687565568538\\
-6	0.922588561284074\\
-5	0.906055554695112\\
-4	0.885243628862301\\
-3	0.859184687293612\\
-2	0.826910966833327\\
-1	0.787553965941664\\
0	0.739847065618625\\
1	0.683488785271801\\
2	0.617922353187391\\
3	0.543611368329271\\
4	0.462774781228502\\
5	0.377985760097794\\
6	0.293669682101864\\
7	0.213981901821892\\
8	0.144462688644623\\
9	0.088604111899831\\
10	0.0483620923865493\\
};

\addplot [line width=0.6mm, color=mycolor4] table[row sep=crcr]{%
-10	0.969766556035403\\
-9	0.964175275957165\\
-8	0.957039379199887\\
-7	0.947784991734834\\
-6	0.935922526780015\\
-5	0.920637611495337\\
-4	0.901087352487056\\
-3	0.876113504187174\\
-2	0.84455635523047\\
-1	0.805416428485289\\
0	0.757382414007667\\
1	0.699899258733029\\
2	0.632487153083888\\
3	0.555795119240353\\
4	0.472380368140137\\
5	0.38482045256584\\
6	0.298077183285548\\
7	0.216594625980236\\
8	0.145841814790528\\
9	0.0892311522708439\\
10	0.0486361641039419\\
};

\definecolor{darkgreen}{rgb}{0.00000,0.39200,0.00000} 

\draw [stealth-,thick, darkgreen] (axis cs:-1.8,0.99) -- (axis cs:-1.5,0.92) node[below]{$R_1=5$ m};
\draw [thick, darkgreen] (-2,0.99) ellipse (0.5cm and 0.2cm); 

\draw [stealth-,thick, darkgreen] (axis cs:-4.5,0.87) -- (axis cs:-3.8,0.7) node[below]{$R_1=35$ m}; 
\draw [thick, darkgreen] (-4.5,0.87) ellipse (0.7cm and 0.7cm); 

\draw [dashed, thick] (axis cs:-1,0.97) rectangle (axis cs:1,1);

\draw[->, thick] (axis cs:0,0.97) -- (3.7in,4.55in);

\end{axis}

\begin{axis}[%
width=1.2in,
height=0.7in,
at={(4in,3in)}, 
scale only axis,
xmin=-1,
xmax=1,
ymin=0.98,
ymax=1,
ytick={0.98, 0.99, 1}, 
axis background/.style={fill=white},
xmajorgrids,
ymajorgrids
]

\addplot [line width=0.6mm, color=mycolor1] table[row sep=crcr]{-1 0.9834\\ 0 0.98098\\ 1 0.97814\\};
\addplot [line width=0.6mm, color=mycolor2] table[row sep=crcr]{-1 0.99082\\ 0 0.98936\\ 1 0.98762\\};
\addplot [line width=0.6mm, color=mycolor3] table[row sep=crcr]{-1 0.99346\\ 0 0.99228\\ 1 0.99088\\};
\addplot [line width=0.6mm, color=mycolor4] table[row sep=crcr]{-1 0.99488\\ 0 0.99386\\ 1 0.99264\\};

\end{axis}

\end{tikzpicture}
}%
     \subfloat[     \label{CLU_SHAPE_PARAMTER}]{%
%
%
\definecolor{mycolor1}{rgb}{0.00000,0.44700,0.74100}%
\definecolor{mycolor2}{rgb}{0.85000,0.32500,0.09800}%
\definecolor{mycolor3}{rgb}{0.92900,0.69400,0.12500}%
\begin{tikzpicture}[scale=0.42, transform shape,font=\Large]
\begin{axis}[%
width=4.521in,
height=3.559in,
at={(0.758in,0.488in)},
scale only axis,
xmin=0,
xmax=0.04,
xtick={0,0.01,0.02,0.03,0.04,0.05},
xticklabels={0,0.01,0.02,0.03,0.04,0.05},
scaled x ticks=false,
xlabel style={font=\huge\color{white!15!black}},
xlabel={$\sigma_{\text{av}_{cl}}$},
ymin=0.62,
ymax=0.75,
ylabel style={font=\Large\color{white!15!black}},
ylabel={Average Sensing Coverage Probability},
axis background/.style={fill=white},
xmajorgrids,
ymajorgrids,
legend style={legend cell align=left, align=left, draw=white!15!black}
]
\addplot [line width=0.6mm, color=mycolor1]
  table[row sep=crcr]{%
0	0.748884\\
0.0111111111111111	0.716996\\
0.0222222222222222	0.683748\\
0.0333333333333333	0.652848\\
0.0444444444444444	0.624284\\
0.0555555555555556	0.593696\\
0.0666666666666667	0.567664\\
0.0777777777777778	0.541256\\
0.0888888888888889	0.516312\\
0.1	0.495036\\
};
\addlegendentry{k=0.7}

\addplot [line width=0.6mm, color=mycolor2]
  table[row sep=crcr]{%
0	0.748892\\
0.0111111111111111	0.714108\\
0.0222222222222222	0.681732\\
0.0333333333333333	0.649712\\
0.0444444444444444	0.617888\\
0.0555555555555556	0.588924\\
0.0666666666666667	0.559668\\
0.0777777777777778	0.535744\\
0.0888888888888889	0.507932\\
0.1	0.483412\\
};
\addlegendentry{k=1 (exponential)}

\addplot [line width=0.6mm, color=mycolor3]
  table[row sep=crcr]{%
0	0.74882\\
0.0111111111111111	0.712184\\
0.0222222222222222	0.677612\\
0.0333333333333333	0.643548\\
0.0444444444444444	0.612592\\
0.0555555555555556	0.58346\\
0.0666666666666667	0.552168\\
0.0777777777777778	0.52378\\
0.0888888888888889	0.497728\\
0.1	0.474172\\
};
\addlegendentry{k=1.5}

\end{axis}

\begin{axis}[%
width=5.833in,
height=4.375in,
at={(0in,0in)},
scale only axis,
xmin=0,
xmax=1,
ymin=0,
ymax=1,
axis line style={draw=none},
ticks=none,
axis x line*=bottom,
axis y line*=left
]
\end{axis}
\end{tikzpicture}%
}%
\caption{(a) The effect of different interference types on the monostatic and the bistatic ($n=2$) operations. (b) The sensing coverage probability for different target locations. (c) The average networked sensing coverage probability with $N=4$ against the scale parameter $\sigma_{\text{av}_{cl}}$ of the Weibull clutter distribution for shape parameters $k \in \{0.7, 1, 1.5\}$.} 
\end{figure*}

Fig.~\ref{eff_sources_mono} depicts the effect of different interference types on monostatic sensing (assuming perfect SIC) and bistatic sensing ($n=2$) for a target positioned 20~m from the serving BS. The close alignment between analytical results and simulations confirms the validity of Theorem 1 and Theorem 2  and supports the statistical independence assumptions used to factor the LT of the aggregate interference into separate components. Moreover, it verifies the accuracy of using the \(R_1\) range to approximate both the target and clutter scatterer ranges within the target’s resolution cell.
 The figure demonstrates that focusing on direct (co-channel) interference, as in communication systems, while ignoring clutter-related interference results in overly optimistic sensing coverage estimates. While all interference types contribute comparably in monostatic sensing, their impact on bistatic sensing varies.
The larger bistatic resolution cell introduces more scatterers, making intra-clutter interference dominant. Meanwhile, inter-clutter interference is lower than in the monostatic case due to independent LoS/NLoS conditions for both inter-clutter links (i.e., $R_n$ and $R_v$) in bistatic scenarios, unlike monostatic sensing, where one inter-clutter link (i.e., $R_1$) is always LoS.
 Overall, monostatic sensing achieves higher coverage probability under effective SIC due to proximity to the target, better LoS conditions, and smaller resolution cells that reduce intra-clutter interference.

Fig.~\ref{dist_effect} illustrates the sensing coverage probability for two scenarios: one with a target near the cell center and the other at a farther location, comparing monostatic sensing with networked sensing involving 2, 3, and 4 BSs. The agreement between the analytical results and the simulations validates~(\ref{joint_fus_form}) for networked sensing fusion and confirms the independence assumption discussed in Remark~\ref{ind_asu_net}. 
The figure shows that dual-mode networked sensing fusion significantly enhances performance by leveraging spatial diversity. This improvement stems from the fact that, while the average coverage probability of individual bistatic links is lower than that of monostatic sensing, some instantaneous bistatic SINRs can exceed monostatic ones due to inherent randomness in the network.
The figure shows that monostatic sensing is sufficient for targets near the cell center. However, as the target becomes closer to the cell edge, the benefits of networked sensing become more evident, providing significant performance enhancement. While involving additional BSs consistently improves coverage probability, the marginal gains diminish as $n$ increases, rendering further cooperation less beneficial when weighed against the practical complexity.

Fig.~\ref{CLU_SHAPE_PARAMTER} illustrates the impact of clutter statistics on networked sensing performance, plotting the simulated coverage probability versus the Weibull scale parameter for shape parameters \(k \in \{0.7, 1, 1.5\}\). Two key insights emerge. First, coverage probability decreases monotonically as clutter strength increases across all distributions, as expected since the mean RCS is directly proportional to the scale parameter. Second, heavier-tailed clutter (\(k=0.7\)) yields higher coverage than lighter-tailed clutter (\(k=1.5\)), with the exponential case (\(k=1\)) in between. This result stems from the spatial averaging inherent in our coverage probability metric over the PPP clutter field. For a fixed scale parameter, decreasing $k$ concentrates substantial probability mass near zero while generating rare extreme values. These abundant small values outweigh the impact of rare extreme spikes, leading to higher coverage probability for smaller $k$. Moreover, the gap widens as $\sigma_{\text{av}_{cl}}$ increases because when clutter is weak, the regime is noise-limited and the shape has little effect; as clutter strengthens, the regime becomes clutter-limited and more sensitive to distributional differences. Overall, the influence of \(k\) is modest and does not change our qualitative conclusions or design recommendations; adopting \(k=1\) remains a tractable baseline that captures the essential trends.

\begin{figure*}
\centering
   \subfloat[  \label{beam_tradeoff}]{%
   \input{beamwidth}
   }
     \subfloat[  \label{bs_avg_tot}]{%
%
%
\definecolor{mycolor1}{rgb}{0.00000,0.44700,0.74100}%
\definecolor{mycolor2}{rgb}{0.85000,0.32500,0.09800}%
\begin{tikzpicture}[scale=0.42, transform shape,font=\Large]

\begin{axis}[%
width=4.521in,
height=3.477in,
at={(0.758in,0.57in)},
scale only axis,
xmin=1,
xmax=500,
xlabel style={font=\Large\color{white!15!black}},
xlabel={$\text{BSs Density/ km}^\text{2}$},
ymin=2,
ymax=41,
ylabel style={font=\Large\color{white!15!black}},
ylabel={Average Rate  (nats/s/Hz)},
axis background/.style={fill=white},
xmajorgrids,
ymajorgrids,
legend style={at={(0.316,0.057)}, anchor=south west, legend cell align=left, align=left, draw=white!15!black}
]

\addplot [line width=0.6mm, color=mycolor1]
  table[row sep=crcr]{%
10	2.37932777777778\\
30.2	5.57799587628866\\
50.4	8.6849\\
70.6	11.5778113989637\\
90.8	13.8997435344828\\
111	15.8910490636704\\
131.2	17.7826355704698\\
151.4	19.4976349693252\\
171.6	20.9657779036827\\
191.8	22.3941276595745\\
212	23.7028492462312\\
232.2	24.8676038277512\\
252.4	26.0530315668203\\
272.7	27.0710825221239\\
292.9	27.9856311965812\\
313.1	28.7089896907216\\
333.3	29.6136453629032\\
353.5	30.4211102362205\\
373.7	30.9881888246628\\
393.9	31.7447516981132\\
414.1	32.3633411111111\\
434.3	32.8437114337568\\
454.5	33.5133944543828\\
474.7	34.1048585840708\\
494.9	34.3071941074523\\
515.102040816327	34.5158857142857\\
535.30612244898	34.9026285714286\\
555.510204081633	35.2832571428571\\
575.714285714286	35.6657714285714\\
595.918367346939	35.9598857142857\\
616.122448979592	36.2878857142857\\
636.326530612245	36.6254857142857\\
656.530612244898	36.8828571428571\\
676.734693877551	37.1316\\
696.938775510204	37.3717142857143\\
717.142857142857	37.5252\\
737.34693877551	37.7229142857143\\
757.551020408163	37.9309714285714\\
777.755102040816	38.1334857142857\\
797.959183673469	38.3063428571429\\
818.163265306122	38.4848\\
838.367346938776	38.6385142857143\\
858.571428571429	38.8000571428571\\
878.775510204082	38.9079428571429\\
898.979591836735	39.0440571428571\\
919.183673469388	39.0713714285714\\
939.387755102041	39.1737714285714\\
959.591836734694	39.2801142857143\\
979.795918367347	39.4028571428572\\
1000	39.542\\
};
\addlegendentry{Dual-mode Networked Sensing}

\addplot [line width=0.6mm, color=mycolor2]
  table[row sep=crcr]{%
10.000000	27.437364\\
30.204082	35.755709\\
50.408163	38.520553\\
70.612245	39.761418\\
90.816327	40.453270\\
111.020408	40.511550\\
131.224490	40.399799\\
151.428571	39.920911\\
171.632653	39.405524\\
191.836735	38.864594\\
212.040816	38.309726\\
232.244898	37.658984\\
252.448980	37.029868\\
272.653061	36.347645\\
292.857143	35.751329\\
313.061224	35.172898\\
333.265306	34.570598\\
353.469388	33.955722\\
373.673469	33.428646\\
393.877551	32.843270\\
414.081633	32.316868\\
434.285714	31.789932\\
454.489796	31.269669\\
474.693878	30.784219\\
494.897959	30.327161\\
515.102041	29.865955\\
535.306122	29.465751\\
555.510204	29.089850\\
575.714286	28.724412\\
595.918367	28.333438\\
616.122449	27.949873\\
636.326531	27.597796\\
656.530612	27.309886\\
676.734694	27.047476\\
696.938776	26.705471\\
717.142857	26.371944\\
737.346939	26.079903\\
757.551020	25.792530\\
777.755102	25.519985\\
797.959184	25.257380\\
818.163265	24.981133\\
838.367347	24.730662\\
858.571429	24.535717\\
878.775510	24.378532\\
898.979592	24.156601\\
919.183673	23.930964\\
939.387755	23.772724\\
959.591837	23.598513\\
979.795918	23.428949\\
1000.000000	23.263942\\
};
\addlegendentry{Communication}

\addplot [
    color=white, 
    draw=none, 
    mark=x, 
    thick, 
    mark size=4pt, 
    mark options={black},
]
  table[row sep=crcr]{%
10	2.23782793142858\\
30.2040816326531	5.59547753142857\\
50.4081632653061	8.62238290285715\\
70.6122448979592	11.3185440457143\\
90.8163265306122	13.6235419885715\\
111.020408163265	15.6592925714285\\
131.224489795918	17.4786127542857\\
151.428571428571	19.1226328228572\\
171.632653061224	20.6394042971428\\
191.836734693878	21.99885952\\
212.040816326531	23.2980092342857\\
232.244897959184	24.51262912\\
252.448979591837	25.5708971885715\\
272.65306122449	26.5821349028572\\
292.857142857143	27.4904618057143\\
313.061224489796	28.3270801828572\\
333.265306122449	29.1070238628572\\
353.469387755102	29.8495248\\
373.673469387755	30.4714714514285\\
393.877551020408	31.1634246857143\\
414.081632653061	31.7772587428572\\
434.285714285714	32.3722579657143\\
454.489795918367	32.9187518171428\\
474.69387755102	33.3865591771428\\
494.897959183673	33.8256604342857\\
515.102040816327	34.2673713371428\\
535.30612244898	34.6513296457143\\
555.510204081633	35.0292176914285\\
575.714285714286	35.4089778742857\\
595.918367346939	35.7009745371428\\
616.122448979592	36.0266129371428\\
636.326530612245	36.3617822171428\\
656.530612244898	36.6173005714285\\
676.734693877551	36.86425248\\
696.938775510204	37.1026379428572\\
717.142857142857	37.25501856\\
737.34693877551	37.4513093028572\\
757.551020408163	37.6578684342857\\
777.755102040816	37.8589246171428\\
797.959183673469	38.0305371885715\\
818.163265306122	38.20770944\\
838.367346938776	38.3603169828572\\
858.571428571429	38.5206967314285\\
878.775510204082	38.6278056685715\\
898.979591836735	38.7629399314285\\
919.183673469388	38.7900575542857\\
939.387755102041	38.8917202742857\\
959.591836734694	38.9972974628572\\
979.795918367347	39.1191565714286\\
1000	39.2572976\\
};
\addlegendentry{Simulations}

\addplot [
    color=white, 
    draw=none, 
    mark=x, 
    thick, 
    mark size=4pt, 
    mark options={black},
]
  table[row sep=crcr]{%
10	27.1459412\\
30.4166666666667	35.235809\\
50.8333333333333	38.1978296\\
71.25	39.3954008\\
91.6666666666667	39.9854392\\
112.083333333333	40.0475642\\
132.5	39.82461\\
152.916666666667	39.5834656\\
173.333333333333	38.953866\\
193.75	38.4322148\\
214.166666666667	37.653217\\
234.583333333333	37.0087074\\
255	36.510614\\
275.416666666667	35.808353\\
295.833333333333	35.220899\\
316.25	34.6107818\\
336.666666666667	34.1191494\\
357.083333333333	33.3917402\\
377.5	32.789078\\
397.916666666667	32.2288596\\
418.333333333333	31.6863344\\
438.75	31.2312812\\
459.166666666667	30.747899\\
479.583333333333	30.2094492\\
500	29.8409734\\
};

\end{axis}

\begin{axis}[%
width=5.833in,
height=4.375in,
at={(0in,0in)},
scale only axis,
xmin=0,
xmax=1,
ymin=0,
ymax=1,
axis line style={draw=none},
ticks=none,
axis x line*=bottom,
axis y line*=left
]
\end{axis}
\end{tikzpicture}%
}%
     \subfloat[    \label{overhead_networked-sensingn}]{%
   \includegraphics[width=0.33\textwidth]{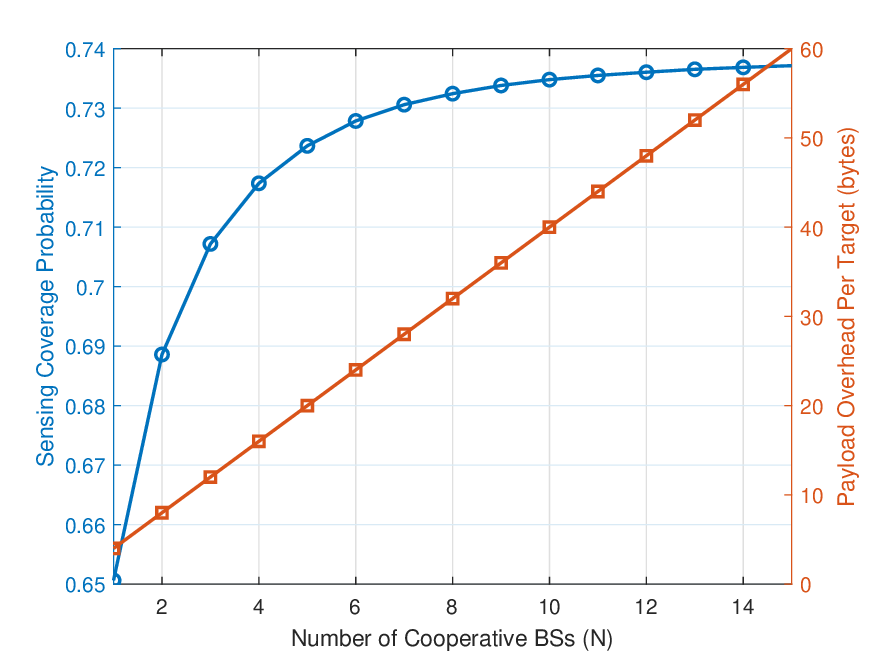}
}%
\caption{ (a)  The average coverage probability versus beamwidth spread. (b) The average sensing and communication rates versus BS density. (c) Average sensing coverage probability (left $y$-axis) and the payload backhaul overhead per target (right $y$-axis) versus the number of cooperating BSs.  }
\end{figure*}

Fig. \ref{beam_tradeoff} shows the average coverage probability for sensing and communication versus 3-dB beamwidth, validating Theorem \ref{com_cov_prob}. For sensing, narrower beams improve coverage probability by increasing antenna gain, reducing direct interference, minimizing clutter, and enhancing angular resolution. In communication, narrower beams theoretically improve coverage due to higher gain and reduced interference. However, in practice, coverage peaks at an optimal beamwidth and declines sharply beyond it, as narrower beams provide higher gain but are more sensitive to misalignment.  Hence, careful beamwidth selection is essential to balance sensing and communication performance in unified ISAC systems.

Fig.~\ref{bs_avg_tot} illustrates the relationship between average sensing and communication rates and BS density, validating Theorems 3 and 5. The plot reveals an optimal BS density for maximizing communication rates. At low densities, the system is noise-limited, where increased density shortens link distances, improving coverage and rates. Beyond this point, the system becomes interference-limited, as higher density transitions more interference links from NLoS to LoS, reducing coverage and rates.
In contrast, sensing rates consistently increase with BS density, growing rapidly at first and slowing at higher densities. This behavior is driven by: 1) increased probability of LoS links in multistatic sensing due to shorter distances;
2) the intended sensing signal experiences path loss with an exponent twice that of the direct interference signal, resulting in a greater marginal improvement in intended signal strength as BS density increases; and 3) the reduction in the resolution cell area, which minimizes the impact of clutter. The slower growth at higher densities is due to more interference signals transitioning to LoS, amplifying their impact, but this effect remains outweighed by the three  aforementioned factors.

Fig.~\ref{overhead_networked-sensingn} illustrates the interplay between cooperative gain and backhaul overhead in networked sensing, where each BS contributes $b_{\text{SC}}=32$ bits (2 bytes for estimate + 2 bytes for SINR) per target. While the backhaul overhead for selection combining grows linearly with the number of cooperating BSs \(N\), the sensing coverage probability exhibits diminishing marginal gains as \(N\) increases, because additional BSs are typically farther from the target and experience poorer channel conditions. For example, the average coverage probability rises only from \(0.727\) at \(N{=}6\) to \(0.736\) at \(N{=}12\), even as the backhaul overhead doubles over the same range.
Since this overhead scales with the number of simultaneous sensed targets, it is essential to cap the cooperation size \(N\) to a reasonable limit. A practical design rule is to increase \(N\) until the incremental coverage improvement falls below a predefined threshold or the total overhead approaches a backhaul budget. Hence, limiting cooperation to small clusters of ($N \approx 4-6$) captures most of the spatial diversity gains while maintaining manageable backhaul requirements.

\begin{figure*}
\centering
\subfloat[ \label{imper_sic}]{%
  \input{imperfect_SIC}
}%
\subfloat[ \label{no_fd}]{%
 \input{No_full_duplex}
}%
\subfloat[  \label{energy_eff}]{%
 \input{Energy_tradeoff}
}%
    \caption{  (a)  The effect of imperfect SIC  for a target at a range = 20 m. (b) The networked-sensing performance without the monostatic sensing in case  FD is not available. (c) The effect of energy allocation between sensing and communication across the ISAC signal.}
\end{figure*}

Fig.~\ref{imper_sic} illustrates the impact of imperfect SIC on sensing performance. The multistatic-only performance for $N=6$ BSs is shown as a horizontal line, since multistatic returns are unaffected by SI. A vertical dashed line marks the intersection point between the monostatic-only and multistatic-only curves, defining two distinct operational regions.
To the left of this threshold (the monostatic-dominant region), monostatic performance exceeds multistatic-only performance. Here, efficient SIC with low residual power enables high sensing coverage primarily through monostatic sensing, while multistatic sensing provides supplementary gains via spatial diversity. To the right of this threshold (the multistatic-dominant region), multistatic performance surpasses monostatic performance. In this regime, monostatic performance declines sharply due to increasing residual SI, and the system becomes sustained primarily by multistatic returns.

This analysis reveals a fundamental design trade-off between SIC capability and coordination complexity. For systems with effective SIC (small $\zeta$), monostatic sensing dominates and minimal coordination suffices. However, as $\zeta$ increases beyond the critical threshold, the system transitions to multistatic-dominant operation, requiring more cooperating BSs to maintain performance at the cost of increased backhaul overhead.
From a system design perspective, this offers a clear choice between two approaches: (1) investing in higher-quality SIC (lower $\zeta$) to preserve monostatic dominance with lighter coordination, or (2) accepting a higher $\zeta$ and compensating by increasing the cooperation level $N$, thereby trading off higher fusion overhead for sustained sensing coverage.
Moreover, the figure shows that targeting \(\approx 100\text{-}120\,\mathrm{dB}\) total suppression is appropriate and aligns with experimental reports in the literature~\cite{barneto2019full}. Further suppression beyond this range yields only marginal gains for our metrics, whereas relaxing suppression below \(\approx 100\,\mathrm{dB}\) substantially degrades monostatic sensing and effectively shifts the system toward reliance on multistatic returns.

Fig. \ref{no_fd} examines average sensing coverage probability without FD, where monostatic sensing is infeasible, relying instead on multistatic sensing. The figure shows that increasing the number of cooperating BSs improves performance by exploiting spatial diversity, as bistatic RCS is captured from multiple angles. The performance also improves with higher network density, driven by the same factors discussed in Fig.~\ref{bs_avg_tot}.  For instance, when six BSs cooperate in receiving sensing echoes instead of a single BS Rx, the sensing coverage probability increases by \textbf{108\%} at a BS density of 250 BSs/km$^\text{2}$ and by \textbf{120\%} at 500 BSs/km$^\text{2}$. This improvement is only possible through the cellular infrastructure’s networking capabilities and synchronized operation, enabling efficient coordination that standalone sensing systems cannot achieve. This seamless cooperation among BSs highlights the significant coordination gain in ISAC networks.

Fig. \ref{energy_eff} analyzes the impact of power allocation on average information rates, where \(E_t\) is chosen so that the average transmit power per time slot equals \(1\) W, i.e., \(E_t/T_t=1~\mathrm{W}\).
The results demonstrate the integration gain of the proposed ISAC system, where the total rate significantly exceeds the individual rates of sensing and communication. This gain is maximized by allocating more energy to the sensing pulse, which helps mitigate SI, compensate for round-trip path loss, and reduce direct interference.
In dense networks, increasing power has a limited effect on communication performance, as it amplifies both the signal and interference, whereas sensing benefits more directly. Additionally, Fig.~\ref{energy_eff} compares the proposed system to a time-sharing approach in which an entire slot is dedicated to sensing, followed by a separate slot for communications. The sensing slot still uses a pulse structure; after transmitting the pulse, the remainder of the slot is used solely for listening and receiving echoes, with no data transmission. Although this eliminates SI, the overall performance is very low, as each function is limited to half the time.

To assess the impact of incorporating sensing functionality on communication, the proposed system is contrasted to a communication-only network, where all resources are allocated to communication. The results show that ISAC maintains comparable communication performance while delivering substantial sensing gains. Notably, with efficient SIC, monostatic sensing increases total throughput by \textbf{90\%}, while dual-mode sensing increases it by \textbf{106\%} relative to a communication-only network, highlighting the ISAC integration gain.
Moreover, Fig. \ref{energy_eff} plots the total ISAC rate without FD, relying on multistatic sensing from the nearest six BSs. Remarkably, the system still achieves a meaningful integration gain, with total throughput increasing by \textbf{60\%} compared to the communication-only network. This demonstrates the proposed ISAC framework’s efficiency even without FD technology. Notably, the optimal power to maximize gain in this scenario is lower than in the FD-enabled cases, as there is no monostatic operation or residual SI power.

\subsection{Trade-off Analysis and Design Guidelines}

Based on our analytical framework and numerical results, we identify several fundamental trade-offs in mmWave ISAC network design and derive the following guidelines:

\begin{itemize}

    \item \textbf{Operation Mode Selection:} System designers face a critical choice between investing in SIC hardware versus coordination complexity. When backhaul capacity is constrained, prioritize enhancing SIC to maintain efficient monostatic operation. Conversely, when hardware constraints limit SIC performance, shift to multistatic operation with moderate cooperation to recover sensing performance.

\item \textbf{Unified ISAC Signal Design:} As demonstrated in Fig.~\ref{energy_eff}, adopt unified ISAC waveform design over time-sharing approaches, as simultaneous operation provides substantial throughput gains even without FD capabilities.

\item \textbf{Cooperation Cluster Sizing:} The number of cooperating BSs should be carefully selected, as sensing coverage exhibits diminishing gains with higher cluster sizes while backhaul overhead grows linearly. Critically, since backhaul capacity is shared between S\&C, excessive sensing cooperation can degrade communication performance by consuming transport resources needed for user data.

\item \textbf{Density-Aware Network Planning:} Select a BS density that balances sensing and communication requirements. As shown in Fig.~\ref{bs_avg_tot}, increasing density beyond the communication-optimal point to improve sensing does not impose a major communication-rate penalty. Thus, a moderate-to-high density can be chosen to favor sensing performance without significantly harming communication.

\item \textbf{Power Allocation Strategy:} To enhance total ISAC throughput, bias power allocation toward sensing as shown in Fig.~\ref{energy_eff}. This is effective because sensing benefits more from increased power (mitigating round-trip loss and SI), while communication in dense networks gains little due to the coupled amplification of desired and interference signals.

\item \textbf{Beamwidth-Aware System Design:} Operate at moderate beamwidths to balance ISAC performance. Avoid ultra-narrow beams that cause communication failure from misalignment and very wide beams that degrade both functions, particularly sensing performance due to reduced gain and resolution.

    \item \textbf{Clutter-Aware System Design:} Incorporate clutter mitigation techniques into ISAC system architecture, as Fig.~\ref{eff_sources_mono} demonstrates that clutter interference significantly impacts sensing performance and cannot be neglected in design.

\end{itemize}

\section{Conclusion and Future Work}\label{con_pp}

The paper studies a large-scale mmWave ISAC network using a dual-mode sensing framework that combines monostatic and multistatic approaches. System-level analysis shows that overlooking sensing-specific interference sources and clutter leads to overestimated performance, with interference affecting monostatic and bistatic sensing differently.
The analytical framework reveals intricate design trade-offs across multiple system parameters. BS density must balance the monotonic improvement of sensing against communication performance, which peaks and then declines in interference-limited regimes. Likewise, beamwidth configuration must reconcile the conflict between sensing resolution and communication reliability. We show that power-allocation strategies favoring sensing maximize total ISAC throughput, while cooperative sensing clusters require careful sizing to balance spatial-diversity gains against backhaul overhead. Moreover, with effective SIC, the system can maintain high performance through monostatic sensing with minimal coordination. However, as residual SI power increases beyond a critical threshold, the system transitions to multistatic-dominant operation, requiring additional cooperating BSs to maintain performance at the cost of increased backhaul overhead.

It is demonstrated that communication performance remains resilient despite the integrated sensing functionality, confirming ISAC's feasibility in dense networks. Moreover, dual-mode cooperative sensing enhances sensing performance by leveraging spatial diversity. Specifically, in scenarios with highly imperfect SIC or without FD capability, the system can still maintain reliable sensing through multistatic operation. This performance is further improved by network densification and by adding more cooperating BSs, highlighting the coordination gains in ISAC systems. Comparisons with communication-only and time-sharing systems demonstrate the superior integration gains of the proposed approach. Notably, even in multistatic mode alone, the system achieves a reasonable integration gain, showing that FD is not a strict requirement for ISAC systems. These findings provide actionable guidance for deploying next-generation cellular networks that simultaneously support high-rate communication and precise environmental sensing.

 A natural extension of this work is to explore unified cooperative architectures, such as cell-free massive MIMO, that simultaneously enhance both communication and sensing through user-centric, network-wide resource allocation and joint signal processing.  Another interesting direction is to extend the framework so that downlink communication, uplink communication, and sensing coexist in the same band. This would require advanced allocation of FD resources to manage the complex interference couplings among the three functions. Extending the environmental model by replacing homogeneous PPP clutter with clustered point processes would also better capture the spatial statistics of specific deployment scenarios. Finally, leveraging machine learning for adaptive resource allocation and interference prediction in such complex environments presents a powerful tool for realizing the full potential of ISAC networks.

\appendices
\section{Proof of Lemma 1 }
Using the definition of LT:
\small
\begin{equation}
\begin{aligned}[b]
\mathcal{L}_{I_{L_s}} (s)&=\mathbb{E}_{\bold\Phi_{L_s},h_{L,i}}\left[\exp\left(-sI_{L_s}\right)\right]\\
&=\mathbb{E}_{\bold\Phi_{L_s},h_{L,i}}\left[\exp\left(-s \sum\limits_{\substack{\text{BS}_i\in \bold\Phi_{L_s}}}   P_c h_{L,i}G_m^2 C_L r_i^{-\eta_L} \right)\right]\\
&=\mathbb{E}_{\bold\Phi_{L_s}}\left[\prod\limits_{\substack{\text{BS}_i\in \bold\Phi_{L_s}}}\mathbb{E}_{h_{L,i}}\left[\exp\left(- s P_c h_{L,i}G_m^2 C_L r_i^{-\eta_L}\right)\right]\right].\\
\end{aligned}
\end{equation}
\normalsize
From the Gamma distribution's moment-generating function:
\small
\begin{equation}
\mathcal{L}_{I_{L_s}} (s)=\mathbb{E}_{\bold\Phi_{L_s}}\left[\prod\limits_{\substack{\text{BS}_i\in \bold\Phi_{L_s}}}\left(1+\frac{ s P_c G_m^2 C_L r_i^{-\eta_L}}{m_L}\right)^{-m_L}\right].
\end{equation}
\normalsize
Consider the closet interfering BS at a distance $R_d$, Using polar coordinates and the definition of probability generating functional (PGFL) in PPP:
\small
\begin{equation}
\begin{aligned}
\mathcal{L}_{I_{L_s}} (s)&=\exp\left(-  \frac{2 \pi \lambda_{BS}}{M^2}    \int_{ R_d}^{\infty}\bold p_{\bold{LOS}}\left(r\right) \right.\\
& \quad \quad \left. \left(1-\left( 1+\frac{s P_c G_m^2 C_L r^{-\eta_L}}{m_L}\right)^{-m_L}    \right) r\;dr  \right).
\end{aligned}
\end{equation}
\normalsize
For NLOS interference, the previous steps are repeated, replacing $\bold p_{\bold{LOS}}\left(r\right)$ by $\left(1-\bold p_{\bold{LOS}}\left(r\right)\right)$, $m_L$ by $m_N$, $C_L$ by $C_N$  and $\eta_L$ by $\eta_N$.
By combining both formulas and taking the expectations over $R_d$ with PDF defined in (\ref{ner_dis}), the lemma is proved.

\section{Proof of Lemma 2}
Since the intra-clutter contribution involves a random number of scatterers and random  RCS within the resolution cell \( A_{rm} \). Therefore, we take the expectation over both the random clutter scatterers and their RCS, then
using the PGFL of PPP:
\small
\begin{equation}
\begin{aligned}
&\mathbb{E}_{cl, \sigma_{cm}} \left[ \exp\left( - \frac{\phi_s \sum_{cl \in \bold\Phi_{cl} \cap A_{rm}} \sigma_{cm}}{\sigma_{\text{av}_t}} \right) \right]\\
&=\mathbb{E}_{cl}\left[ \prod_{cl \in \bold\Phi_{cl} \cap A_{rm}} \mathbb{E}_{\sigma_{cm}}\left[ \exp\left( - \frac{\phi_s  \sigma_{cm}}{\sigma_{\text{av}_t}} \right)\right] \right]\\
&= \exp\left( - \lambda_{cl} \int_{A_{rm}} \mathbb{E}_{\sigma_{cm}}\left[ 1 - \exp\left( - \frac{\phi_s  \sigma_{cm}}{\sigma_{\text{av}_t}} \right) \right] dA_{rm} \right).
\end{aligned}
\end{equation}
\normalsize
By substituting the monostatic range resolution cell area:
\small
\begin{equation}\label{fr_intra}
\begin{aligned}
&\mathbb{E}_{cl, \sigma_{cm}} \left[ \exp\left( - \frac{\phi_s \sum_{cl \in \bold\Phi_{cl} \cap A_{rm}} \sigma_{cm}}{\sigma_{\text{av}_t}} \right) \right] \\
&= \exp \left( - \lambda_{cl} \frac{c \theta_B R_1}{2 W_b} \mathbb{E}_{\sigma_{cm}} \left[ 1 - \exp\left( - \frac{\phi_s  \sigma_{cm}}{\sigma_{\text{av}_t}} \right) \right] \right).
\end{aligned}
\end{equation}
\normalsize
Since \( \sigma_{cm} \) follows a Weibull distribution given by (\ref{clu_wei}) with shape parameter \( k= 1 \) which is exponential, then the expectation simplifies to:
\small
\begin{equation}\label{ff_intra}
\mathbb{E}_{\sigma_{cm}} \left[ 1 - \exp\left( - \frac{\phi_s  \sigma_{cm}}{\sigma_{\text{av}_t}} \right) \right] = \frac{\phi_s  \sigma_{\text{av}_{cl}}}{\sigma_{\text{av}_t} + \phi_s  \sigma_{\text{av}_{cl}}}.
\end{equation}
\normalsize
Thus, by substituting (\ref{ff_intra}) in (\ref{fr_intra}), the lemma is proved.

\section{Proof of Lemma 3}
\small
\begin{equation}
\begin{aligned}[b]
\!\!\!\!\mathcal{L}_{I_{IC1}} (s)&=\mathbb{E}_{\bold\Phi_{L_{IC}},\sigma_{tm}}\left[\exp\left(-sI_{IC1}\right)\right]\\
\!\!\!\!&=\mathbb{E}_{\bold\Phi_{L_{IC}},\sigma_{tm}}\left[\exp\left(-s \sum\limits_{\substack{\text{BS}_{n }\in \bold\Phi_{L_{IC}}\\ n \neq 1}} \cos{\left(\frac{\beta}{2}\right)} R_n^{-\eta_L} \sigma_{tm} \right)\right]\\
\!\!\!\!&=\mathbb{E}_{\bold\Phi_{L_{IC}}}\left[\prod\limits_{\substack{\text{BS}_{n }\in \bold\Phi_{L_{IC}}\\ n \neq 1}}\mathbb{E}_{\sigma_{tm}}\left[\exp\left(- s  \cos{\left(\frac{\beta}{2}\right)} R_n^{-\eta_L} \sigma_{tm}\right)\right]\right].\\
\end{aligned}
\end{equation}
\normalsize
Since \( \sigma_{tm} \) is an exponential PDF with mean \( \sigma_{\text{av}_t} \), then:
\small
\begin{equation}
\mathcal{L}_{I_{IC1}}(s) = \mathbb{E}_{\bold\Phi_{L_{IC}}} \left[ \prod_{\text{BS}_n \in \bold\Phi_{L_{IC}}, n \neq 1} \frac{1}{1 + s \cos\left( \frac{\beta}{2} \right) R_n^{-\eta_L} \sigma_{\text{av}_t}} \right].
\end{equation}
\normalsize
Next, we apply the PGFL of PPP:
\small
\begin{equation}
\begin{aligned}
\mathcal{L}_{I_{IC1}}(s) &= \exp\bigg( - 2 \pi \lambda_{BS} \times \frac{1}{M} \int_{R_{IC}}^{\infty} \bold p_{\bold{LOS}}\left(r\right)\\
&\quad \times \bigg( 1 - \frac{1}{1 + s \cos\left( \frac{\beta}{2} \right) r^{-\eta_L} \sigma_{\text{av}_t}} \bigg) r \, dr \bigg).
\end{aligned}
\end{equation}
\normalsize
Similarly, $\mathcal{L}_{I_{IC2}} (s)$ can be expressed as:
\small
\begin{equation}
\begin{aligned}[b]\label{int_clu_inmd}
\mathcal{L}_{I_{IC2}} (s) 
&= \mathbb{E}_{\boldsymbol{\Phi}_{L_{IC}}} 
\Bigg[ \prod\limits_{\substack{\text{BS}_{n} \in \boldsymbol{\Phi}_{L_{IC}} \\ n \neq 1}} 
\mathbb{E}_{cl, \sigma_{cm}} 
\Bigg[ \exp\bigg( 
-s \cos{\left(\frac{\beta}{2}\right)} R_n^{-\eta_L} \\
&\quad \cdot 
\sum\limits_{cl \in \boldsymbol{\Phi}_{cl} \cap A_{rm}} 
\sigma_{cm} 
\bigg) \Bigg] \Bigg].
\end{aligned}
\end{equation}
\normalsize
As in the proof of Lemma 2, we apply the PGFL for the clutter scatterers and multiply by the area of the resolution cell:
\small
\begin{equation}
\begin{aligned}[b]
\!\!\!&\mathbb{E}_{cl, \sigma_{cm}} \left[ \exp \left( - s \cos \left( \frac{\beta}{2} \right) R_n^{-\eta_L} \sum_{cl \in \bold\Phi_{cl} \cap A_{rm}} \sigma_{cm} \right) \right]\\
\!\!\!&= \exp \left( - \lambda_{cl} \cdot \frac{c \theta_B R_1}{2 W_b} \cdot \mathbb{E}_{\sigma_{cm}} \left[ 1 - \exp \left( - s \cos \left( \frac{\beta}{2} \right) R_n^{-\eta_L} \sigma_{cm} \right) \right] \right).
\end{aligned}
\end{equation}
\normalsize
By using the  the Weibull  distribution when $k = 1$ and compute the expectation over $ \sigma_{cm} $:
\small
\begin{equation}
\!\!\!\!\mathbb{E}_{\sigma_{cm}} \left[ 1 - \exp \left( - s \cos \left( \frac{\beta}{2} \right) R_n^{-\eta_L} \sigma_{cm} \right) \right] = \frac{s \cos \left( \frac{\beta}{2} \right) R_n^{-\eta_L} \sigma_{\text{av}_{cl}}}{1 + s \cos \left( \frac{\beta}{2} \right) R_n^{-\eta_L} \sigma_{\text{av}_{cl}}}.
\end{equation}
\normalsize
We substitute back into (\ref{int_clu_inmd}), then apply the PGFL for the BSs:
\small
\begin{equation}
\begin{aligned}
&\mathcal{L}_{I_{IC2}}(s) = \exp \bigg( - \int_{R_{IC}}^\infty \lambda_{BS} \cdot \frac{1}{M} \cdot \boldsymbol{p}_{\text{LOS}}\big(r\big) \\
&\cdot \bigg( 1 - \exp \bigg( - \lambda_{cl} \cdot \frac{c \theta_B R_1}{2 W_b} \cdot \frac{s \cos \bigg( \frac{\beta}{2} \bigg) r^{-\eta_L} \sigma_{\text{av}_{cl}}}{1 + s \cos \bigg( \frac{\beta}{2} \bigg) r^{-\eta_L} \sigma_{\text{av}_{cl}}} \bigg) \bigg) 2 \pi r \, dr \bigg).
\end{aligned}
\end{equation}
\normalsize
Since $R_{IC}$ and $\beta$ are two RVs with PDFs given by (\ref{cond_dstt}) when $n=2$ and (\ref{bet_distrb}) respectively,
then, the lemma is proved by combining both formulas and taking the expectations.

\section{Proof of Theorem 1}
By simplifying (\ref{scnr_mono}) and substituting $\sigma_{tb}=\sigma_{tm} \cos{\left(\frac{\beta}{2}\right)}$ and $\sigma_{cb}=\sigma_{cm}\cos{\left(\frac{\beta}{2}\right)}$, and by letting 
$L_{IC1}= \sum\limits_{\substack{\text{BS}_{n }\in \bold\Phi_{L_{IC}}\\ n \neq 1}} \cos{\left(\frac{\beta}{2}\right)} R_n^{-\eta_L} \sigma_{tm}$  and  $L_{IC2}= \sum\limits_{\substack{\text{BS}_{n }\in \bold\Phi_{L_{IC}}\\ n \neq 1}} \cos{\left(\frac{\beta}{2}\right)} R_n^{-\eta_L} \sum\limits_{cl\in\bold\Phi_{cl}\cap A_{rm}} \sigma_{cm}$, then 
the monostatic sensing coverage probability can be expressed as:
\small
\begin{equation}
\begin{aligned}[b]
\mathcal{P}_M &= \mathbb{P} \left( \text{SINR}_M > \phi_s \right) \\
&= \mathbb{P} \bigg[\sigma_{tm} > \sum\limits_{cl \in \boldsymbol{\Phi}_{cl} \cap A_{rm}} \phi_s \, \sigma_{cm} + \phi_s \, R_1^{\eta_L} L_{IC1}\quad + \\
& \phi_s \, R_1^{\eta_L} L_{IC2} 
+ \frac{\phi_s \, (4\pi)^3 R_{1}^{2 \eta_L}}{P_s G_{m}^2 \lambda^2} 
\bigg( K_B T W_b + I_{L_s} + I_{N_s} + P_c \zeta \bigg) \bigg].
\end{aligned}
\end{equation}
\normalsize
 Using the Swerling I model, with PDF given by (\ref{sw1_rcs}), Then:
\small
\begin{equation}
\begin{aligned}[b]
&\mathcal{P}_M = \mathbb{E} \Bigg[ \exp\bigg( - \frac{1}{\sigma_{\text{av}_t}} \bigg[\sum_{cl \in \boldsymbol{\Phi}_{cl} \cap A_{rm}} \phi_s \, \sigma_{cm} 
+ \phi_s \, R_1^{\eta_L} L_{IC1} \\
& + \phi_s \, R_1^{\eta_L} L_{IC2} 
+ \frac{\phi_s \, (4\pi)^3 R_1^{2 \eta_L}}{P_s G_m^2 \lambda^2} 
\bigg(K_B T W_b + I_{L_s} + I_{N_s} + P_c \zeta \bigg) \bigg] \bigg) \Bigg].
\end{aligned}
\end{equation}
\normalsize

Finally, owing to the independence assumption of different interference terms, the expression can be decomposed into terms representing the LT of different interference sources.

\section{Proof of Theorem 2}

Considering the target in LoS conditions with the  Rx, then by simplifying (\ref{scnr_bi}) and substituting $\sigma_{tb}=\sigma_{tm} \cos{\left(\frac{\beta}{2}\right)}$ and $\sigma_{cb}=\sigma_{cm}\cos{\left(\frac{\beta}{2}\right)}$, and by letting 
$L_{IC1_b}= \sum\limits_{\substack{\text{BS}_{v}\in \bold\Phi_{L_{IC}}\\ v \neq 1, v \neq n}} \cos{\left(\frac{\beta_I}{2}\right)} R_v^{-\eta_L} \sigma_{tm}$  and  $L_{IC2_b}= \sum\limits_{\substack{\text{BS}_{v}\in \bold\Phi_{L_{IC}}\\ v \neq 1, v \neq n}} \cos{\left(\frac{\beta_I}{2}\right)} R_v^{-\eta_L} \sum\limits_{cl\in\bold\Phi_{cl}\cap A_{rb}} \sigma_{cm}$, then
the bistatic sensing coverage probability can be expressed as:
\small
\begin{equation}\label{rad_cov1_b}
\begin{aligned}[b]
&\mathbb{P} \left(\text{SINR}_{B_n} > \phi_s \right) 
= \mathbb{P} \bigg[\sigma_{tm} > \sum\limits_{cl \in \boldsymbol{\Phi}_{cl} \cap A_{rb}} \phi_s \, \sigma_{cm} 
+ \frac{\phi_s \, R_1^{\eta_L} L_{IC1_b}}{\cos{\left(\frac{\beta}{2}\right)}} \\
&+ \frac{\phi_s \, R_1^{\eta_L} L_{IC2_b}}{\cos{\left(\frac{\beta}{2}\right)}} + \frac{\phi_s \, (4\pi)^3 R_1^{\eta_L} R_n^{\eta_L}}{P_s G_{m}^2 \lambda^2 \cos{\left(\frac{\beta}{2}\right)}} 
\big(K_B T W_b + I_{L_s} + I_{N_s} \big) \bigg].
\end{aligned}
\end{equation}
\normalsize
Again, by adopting the Swerling I model, then:
\footnotesize
\begin{equation}
\begin{aligned}[b]
\!\!\!&\mathbb{P} \left(\text{SINR}_{B_n} > \phi_s \right) 
=  \mathbb{E} \Bigg[ \exp\bigg( - \frac{1}{\sigma_{\text{av}_t}} \bigg[ \sum\limits_{cl \in \boldsymbol{\Phi}_{cl} \cap A_{rb}} \phi_s \, \sigma_{cm} 
+ \frac{\phi_s \, R_1^{\eta_L} L_{IC1_b}}{\cos{\left(\frac{\beta}{2}\right)}} \\
\!\!\!&\quad + \frac{\phi_s \, R_1^{\eta_L} L_{IC2_b}}{\cos{\left(\frac{\beta}{2}\right)}} 
+ \frac{\phi_s \, (4\pi)^3 R_1^{\eta_L} R_n^{\eta_L}}{P_s G_{m}^2 \lambda^2 \cos{\left(\frac{\beta}{2}\right)}} 
\big(K_B T W_b + I_{L_s} + I_{N_s} \big) \bigg] \bigg) \Bigg].
\end{aligned}
\end{equation}
\normalsize
By incorporating the probability that the target is LoS with the RX and that the bistatic return occurs on a beam other than the beam using the same frequency, and owing to the independence of interference terms, the expression can be decomposed into terms representing the LT of different interference sources.
Finally, since $R_n$ and $\beta$ are RVs with PDFs specified in (\ref{cond_dstt}) and (\ref{bet_distrb}), the theorem is proved by taking their expectations.

\section{Proof of Lemma 7}

The proof proceeds similarly to that of Lemma 1, utilizing the definition of LT and the moment-generating function of the Gamma distribution, we can reach:
\small
\begin{equation}
\mathcal{L}_{I_{L_c}} (s)=\mathbb{E}_{\bold\Phi_{L_c}}\left[\prod\limits_{\substack{\text{BS}_i\in \phi_{L_c}}}\left(1+\frac{  s P_c G (\theta_i)C_L r_i^{-\eta_L}}{m_L}\right)^{-m_L}\right].
\end{equation}
\normalsize
Combining BSs' location with their orientations and from the definition of PGFL in PPP, then:
\small
\begin{equation}
\begin{aligned}
\mathcal{L}_{I_{L_c}} (s)&=\exp\left(- \lambda_{BS} \int_{-\frac{\pi}{d}}^{\frac{\pi}{d}}\int_{R_o}^{\infty}\bold p_{\bold{LOS}}\left(r\right) \right.\\ 
&  \left.\left(1-\left( 1+\frac{s P_c G(\theta_i)C_L r^{-\eta_L}}{m_L}\right)^{-m_L}    \right) r\;dr \;d\theta_i \right).
 \end{aligned}
\end{equation}
\normalsize
For NLOS interference, the same steps are followed using the NLOS parameters.

\section{Proof of Theorem 4}
\small
\begin{equation}
\begin{aligned}[b]
&\mathcal{P}_{\text{com}}(R_o)=\mathbb{P} \left(\rm{SINR_C}>\phi_c\right) =\mathbb{P}\left(\frac{P_c h_{L,o}G (\theta_m) C_L R_o^{-\eta_L}}{I_{L_c}+I_{N_c}+K_B T W_b}>\phi_c \right)\\
&=\mathbb{P}\left(h_{L,o}>\phi_c  R_o^{\eta_L}\left(P_c G (\theta_m) C_L\right)^{-1}
\times(I_{L_c}+I_{N_c}+K_B T W_b)\right).
\end{aligned}
\end{equation}
\normalsize
Utilizing Alzer’s inequality \cite{alzer1997some}:
\small
\begin{equation}
\begin{aligned}[b]
\mathcal{P}_{\text{com}}(R_o) \approx & \sum_{n=1}^{m_L} \left(-1\right)^{n+1} {m_L \choose n}\\
&\mathbb{E}\left[\exp\left(-\;\frac{k_L\;n \;\phi_c \; R_o^{\eta_L}\; (I_{L_c}+I_{N_c}+K_B T W_b)}{P_c G (\theta_m) C_L}\right)\right],
\end{aligned}
\end{equation}
\normalsize
where $k_L=m_L(m_L!)^{-\;\frac{1}{m_L}}$. Moreover,
 from the definition of LT, and since \(\boldsymbol{\Phi}_{L_c}\) and \(\boldsymbol{\Phi}_{N_c}\) are independent, then:
\small
\begin{equation}
\begin{aligned}[b]
\mathcal{P}_{\text{com}}(R_o) &=\sum_{n=1}^{m_L} \left(-1\right)^{n+1} {m_L \choose n} \exp\left(-\;\frac{k_L\;n \;\phi_c \; R_o^{\eta_L}\; K_B T W_b}{P_c G (\theta_m) C_L}\right)\\
& \mathcal{L}_{I_{L_c}} \left(\frac{k_L\;n \;\phi_c \; R_o^{\eta_L}}{P_c G (\theta_m) C_L}\right) \mathcal{L}_{I_{N_c}} \left(\frac{k_L\;n \;\phi_c \; R_o^{\eta_L}}{P_c G (\theta_m) C_L}\right).
\end{aligned}
\end{equation}
\normalsize
since $G (\theta_m)$ is a random parameter, let $G (\theta_m)= G_c$ with PDF $f\left(G_c\right)$ then:
\small
\begin{equation}
\begin{aligned}[b]
\mathcal{P}_{\text{com}}(R_o)&=\sum_{n=1}^{m_L} \left(-1\right)^{n+1} {m_L \choose n} \int \exp\left(-\;\frac{k_L\;n \;\phi_c \; R_o^{\eta_L}\; K_B T W_b}{P_c G_c C_L}\right)\\
&\mathcal{L}_{I_{L_c}} \left(\frac{k_L\;n \;\phi_c \; R_o^{\eta_L}}{P_c G_c C_L}\right) \mathcal{L}_{I_{N_c}} \left(\frac{k_L\;n \;\phi_c \; R_o^{\eta_L}}{P_c G_c C_L}\right)f\left(G_c\right) dG_c.
\end{aligned}
\end{equation}
\normalsize   
Finally, to find the average coverage probability, we take the expectation over $R_o$ whose PDF is given by (\ref{ner_dis}).

\bibliographystyle{ieeetr}
\bibliography{bibliography.bib}

\end{document}